\documentclass[aps,rmp,longbibliography,showpacs,twocolumn]{revtex4-1}
\usepackage[dvips]{epsfig}
\usepackage{amsmath}
\usepackage{amsfonts}
\usepackage{amssymb}
\usepackage{graphicx}
\usepackage[usenames]{color}

\begin{document}

\title{Green Algae as Model Organisms for Biological Fluid Dynamics}\thanks{Annual Review of Fluid Mechanics {\bf 47}, in press (2015)}
\author{Raymond E. Goldstein}
\email{R.E.Goldstein@damtp.cam.ac.uk}
\affiliation{Department of Applied Mathematics and Theoretical Physics, Centre for Mathematical
Sciences, University of Cambridge, Wilberforce Road, Cambridge CB3 0WA, United Kingdom}
\date{\today}

\begin{abstract}
In the past decade the volvocine green algae, spanning from the unicellular {\it Chlamydomonas} to
multicellular {\it Volvox}, have emerged as model organisms for a number of problems in
biological fluid dynamics. These include flagellar propulsion, nutrient uptake by swimming organisms, hydrodynamic interactions 
mediated by walls, collective dynamics and transport within
suspensions of microswimmers, the mechanism of phototaxis, and 
the stochastic dynamics of
flagellar synchronization. Green algae are well suited to the study of such problems because of their range of sizes (from 10 $\mu$m 
to several millimetres), their geometric regularity, the ease with which they
can be cultured and the availability of many mutants that allow for connections between molecular details and organism-level behavior. 
This review summarizes
these recent developments and highlights promising future directions in the study of biological fluid
dynamics, especially in the context of evolutionary biology, that can take advantage of these remarkable organisms.
\end{abstract}

\maketitle

\section{VOLVOCINE ALGAE AND THE EVOLUTION OF MULTICELLULARITY}

High on any list of fundamental questions in biology -- a list surely led by the origin of life and the nature of consciousness --
is one regarding evolutionary biology: What are the origins of multicellularity \cite{Bonner98}?  That is,  why and 
how did the simplest unicellular life forms first appearing on Earth evolve into multicellular organisms, which were 
not only larger but eventually exhibited cellular differentiation?  In other words, what are the (fitness) advantages of 
being larger and of dividing up life's processes into specialized cells, increasing complexity 
\cite{Smith,szathmarysmith95}?  For microscopic life existing in a fluid environment, 
these questions unsurprisingly often involve physical processes such as transport (diffusion, mixing, and buoyancy), locomotion, and 
sensing, because the exchange of materials with the environment is one of the most basic features of life.   For these reasons, 
physical scientists are beginning to move toward these problems in evolutionary biology with the hope of shedding 
light on these deep issues.  As with all fundamental problems in biology, there are model organisms
which by community consensus become the focus of research from many differing perspectives.  Examples include
the enteric bacterium {\it Escherichia coli} for sensing and locomotion, the fruit fly {\it Drosophila melanogaster} for genetics and development,
and {\it Arabidopsis thaliana} for much of plant science.  The purpose here is to discuss a class of organisms that
has emerged as a model for many aspects of biological fluid dynamics.  {\it Caveat emptor}: many of the problems discussed in
this review can only be fully appreciated in their native biological context.  Their ultimate solutions will most likely not
involve just the application of fluid dynamics.  Rather, they will require an interdisciplinary approach in which cell biology, genetics, 
microscopy, micromanipulation, and some applied mathematics together will unravel the underlying biological complexity.
Fluid dynamics will have an important role, but the broader scientific method will be key.

\begin{figure}[t]
    \begin{center}
        \includegraphics*[clip=true, width = 0.95\columnwidth]{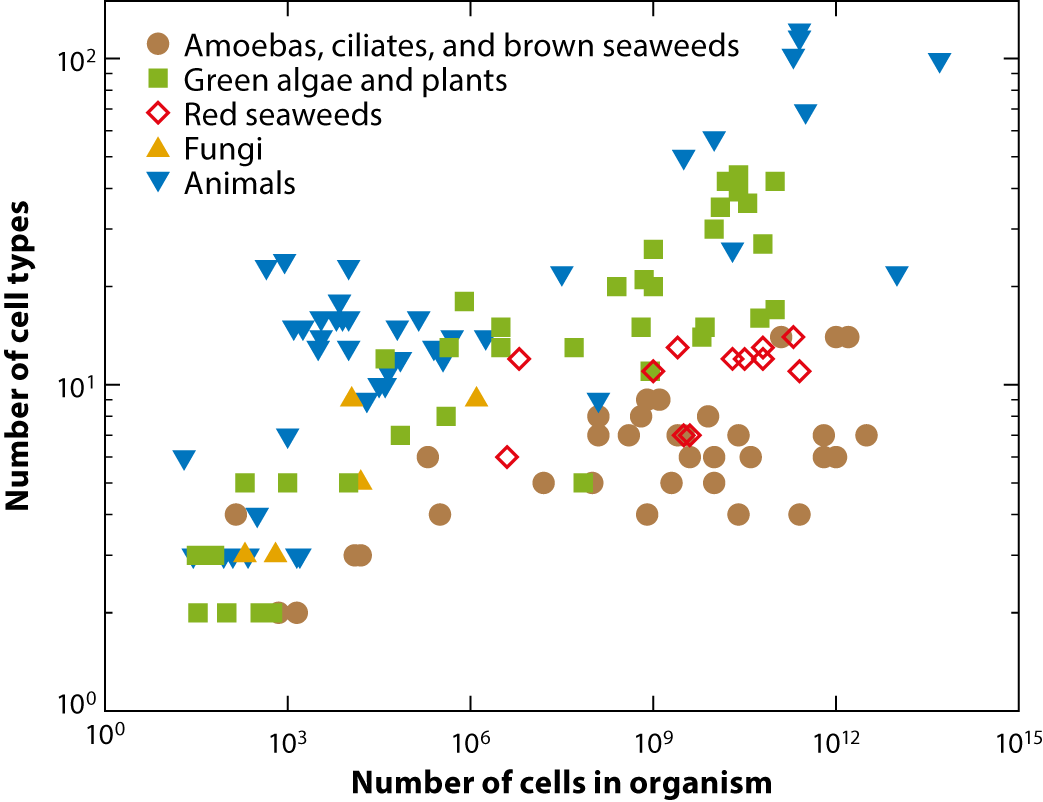}
    \end{center}
    \caption{The number of distinct cell types as a function of cells in an organism's body. Figure adapted from Bell \& Mooers (1997).}
    \label{fig1}
\end{figure}

To put the problem of differentiation, complexity, and size into perspective, let us consider the data shown in Figure \ref{fig1}, which 
were first presented by \textcite{BellMooers} and then revisited by \textcite{Bonner}.  For a range
of organisms in the animal and plant world, this figure shows a double logarithmic plot of the number of distinct cell types found in an individual 
organism as a function of the total number of cells in the organism's body.  Although debate exists about the precise way to count distinct cell 
types in higher organisms, such subtleties are not important for the present, qualitative discussion. The archetypal unicellular organism,
such as a bacterium, is located at $(1,1)$, whereas a human is just off the upper right of the graph at $(10^{14},210)$.  
(The figure $10^{14}$ can be rationalized by the volume of a human, $\sim 0.1$ m$^3$, and the linear dimension of
a typical cell of $\sim 10$ $\mu$m).  There is a 
{\it general} trend that larger organisms are more complex, as measured by the number of cell types,
but the data are noisy, and it would be unwise to try to find a meaningful power law.  Rather, various arguments suggest that
a good way to tackle the size-complexity relationship is to be very reductionist and 
begin at the lower left-hand corner of the graph, where 
organisms of ever-increasing cell number abruptly make the transition from one to two cell types.  

\begin{figure}[b]
    \begin{center}
        \includegraphics*[clip=true, width = 0.95\columnwidth]{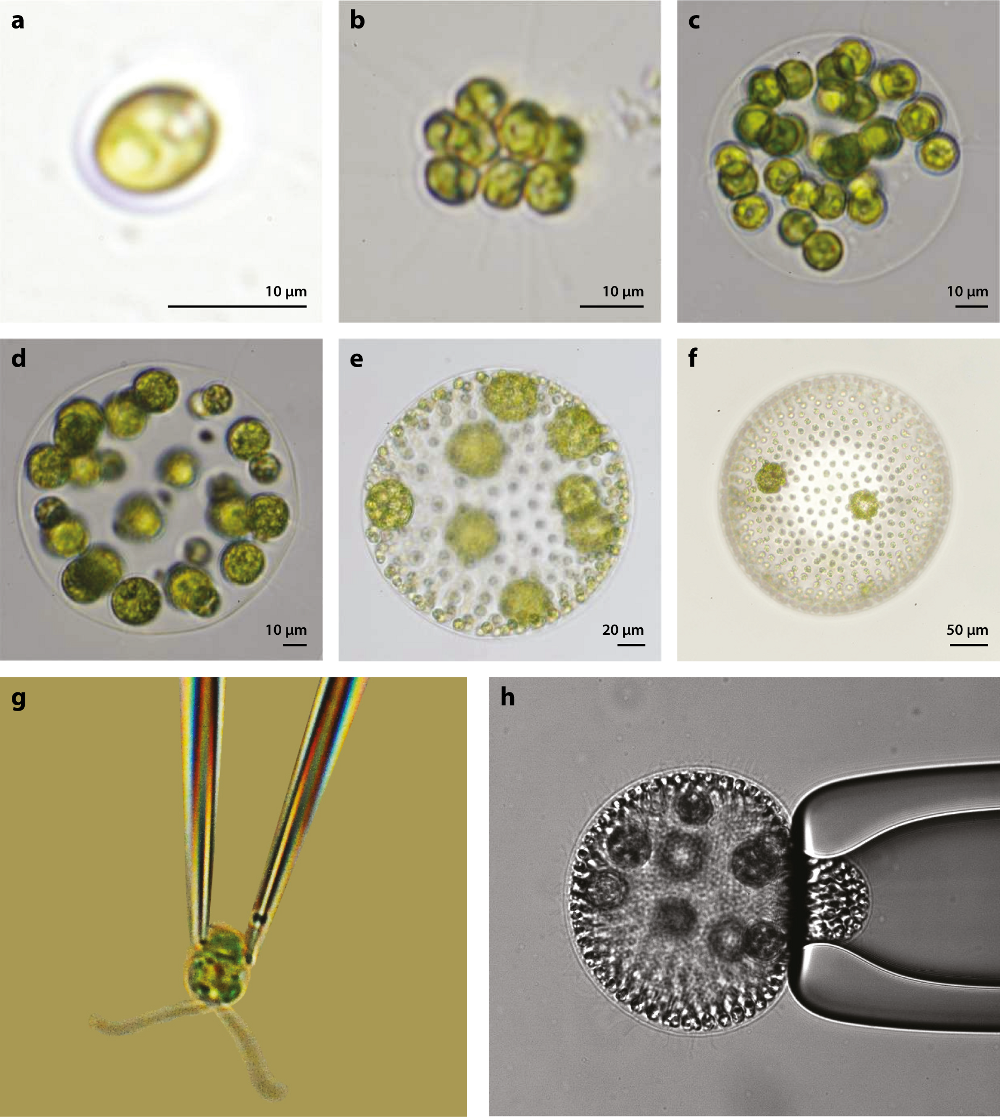}
    \end{center}
    \caption{Examples of volvocine green algae: ({\it a}) unicellular {\it Chlamydomonas reinhardtii}, 
undifferentiated ({\it b}) {\it Gonium pectorale} ($8$ cells) 
and ({\it c}) {\it Eudorina elegans} ($32$ cells), ({\it d}) soma-differentiated {\it Pleodorina californica} ($64$ cells), and 
germ-soma differentiated ({\it e}){\it Volvox carteri} 
($\sim 1,000$ cells) and ({\it f}) {\it Volvox aureus} ($\sim 2,000$ cells). ({\it g}) {\it C. reinhardtii} and ({\it h}) {\it V. carteri} 
held on glass micropipettes. 
Panels a--f reproduced with permission from Solari et al. (2006a). Copyright 2006 by the National Academy of Sciences.}
    \label{fig2}
\end{figure}

In the late 1800s the great biologist August \textcite{Weismann92} suggested volvocine green algae as a class of model organisms for the study of
the transition to multicellularity (Figure \ref{fig2}).  Weismann was the 
originator of germ plasm theory, 
according to which the germ cells ({\it gametes}, from the Greek root meaning `spouse') alone control heredity, 
whereas the sterile somatic cells  (meaning `of the body') in multicellular 
organisms play no role.   This distinction is clearly seen in the volvocine algae, which derive from the unicellular 
biflagellate {\it Chlamydomonas} \cite{Sourcebook}, and continues through species comprising $\sim 2^n$ ($n\lesssim 16$) 
{\it Chlamydomonas}-like cells 
arranged in regular geometric structures, all the way 
to {\it Volvox}, of which there are species containing up to 
$50,000$ cells in a spheroid several millimeters in diameter \cite{Kirkbook}.
This figure shows a sharp divide at {\it Volvox},
for all species shown in Figure 2a-d have cells of only one type that carry out all the functions of life, reproduction and everything 
else, whereas the species shown in Figure 2e,f have two cell types: sterile somatic cells (galley slaves) mounted in a transparent gel-like
extracellular matrix (ECM) with their flagella pointing outwards, and specialized reproductive (germ) cells on the inside. 
This so-called germ-soma differentiation
is the most basic distinction of cell types in biology and can be seen as a first step towards multicellularity.  
A natural question, returned to below, is why nature put the dividing line at {\it Volvox}:
Is there something special about the size of {\it Volvox} or its cell number that suddenly makes it advantageous to have two cell types?

As summarized by \textcite{Kirkbook} in his celebrated book on {\it Volvox}, there are many reasons why this class 
of organisms is excellent for the study of multicellularity; it is an extant lineage spanning from unicellular to 
differentiated multicellular species, it is readily obtainable in nature, it has been studied from a variety of different 
perspectives (biochemical, developmental, genetic), its ecological niches are understood, it developed recently enough 
that its genome may retain traces of genetic changes in organization, it displays evidence of repeated genetic changes,
and it is amenable to DNA transformation.  Indeed, the genome of {\it Volvox} has recently been sequenced 
\cite{Prochnik} and, in comparison
with that of {\it Chlamydomonas}, is providing fascinating information about the genetic changes necessary for 
multicellularity.  For fluid dynamicists theory becomes simpler because these colonies are quasi-spherical 
and their large size makes visualization easy.   In addition, 
because their size spans from $10$ $\mu$m to several millimeters, it may
be possible to use these organisms to uncover scaling laws of physiology and behavior that underlie the transition to 
multicellularity \cite{Multicellular,Flagflows,Motility,Amnat}.

\begin{figure}[t]
    \begin{center}
        \includegraphics*[clip=true, width = 0.95\columnwidth]{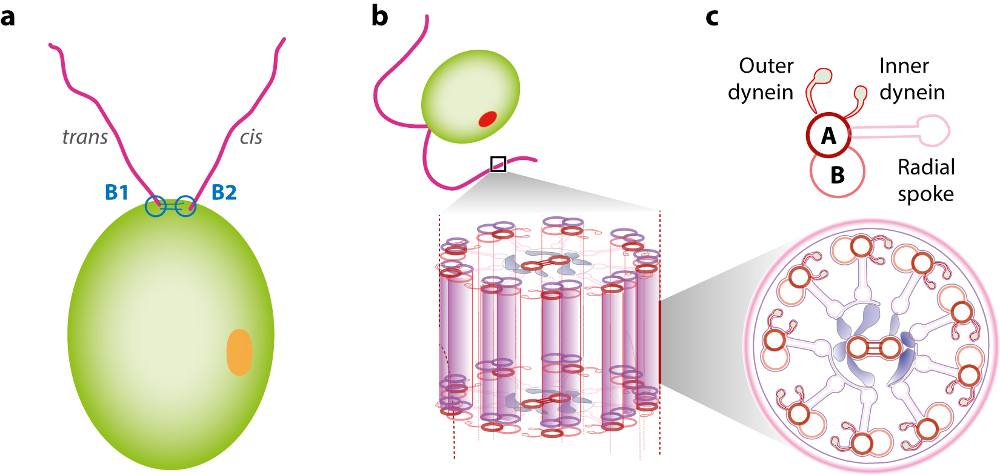}
    \end{center}
    \caption{{\it Chlamydomonas} and the structure of the eukaryotic axoneme. ({\it a}) The two flagella are termed {\it cis} and 
{\it trans} based on their 
position relative to the eyespot (\textit{orange}). ({\it b}) The axoneme consists a $9 + 2$ arrangement of microtubule doublets on which 
dynein motors 
initiate interdoublet sliding and produce the flagellar beat. Figure adapted from Wan et al. (2014).}
    \label{fig3}
\end{figure}

{\it Volvox} itself was discovered by Antony \textcite{vanLeeuwenhoek}, as described in a famous letter to the Royal Society.  
Its name was given by \citet{Linneaus}, in the last entry of his great book on 
taxonomy, based on the Latin word {\it volvere} (to roll) for its characteristic spinning motion 
around a body-fixed axis.    That spinning is a consequence of the coordinated action of the thousands of flagella on its
surface, each of which beat in a plane that is tilted approximately $15^{\circ}$ from the anterior-posterior axis.  All known
species rotate in the same direction (clockwise when viewed from the anterior), but there are known inversion mutants 
that rotate in the opposite sense.   These eukaryotic flagella, often called cilia when
occurring in large numbers (as on protists such as {\it Paramecium} or {\it Opalina}) or when emanating from tissues
within an organism (as in human respiratory and reproductive systems), must be distinguished from the 
{\it prokaryotic} flagella of bacteria.  The latter are rigid, generally helical structures comprising the protein 
flagellin, turned by the proton-driven motion of rotary 
motors embedded in the cell wall.  Eukaryotic flagella and cilia have a much more complex microstructure 
(Figure \ref{fig3}).  The basic structure, known as the axoneme, consists of parallel microtubules (themselves
polymerized tubulin protein monomers) arranged in the form of nine doublets around the periphery and two doublets in the center.
This 9+2 arrangement is one of the most highly conserved structures in biology, such that the protein content
in {\it Chlamydomonas} flagella, which appeared on Earth the better part of a billion years ago, is essentially identical
to that in human cilia \cite{Pazour}.  For this reason, the volvocine algae, and {\it Chlamydomonas} in 
particular, have long been model organisms in the study of human ciliopathies \cite{IbanezTallon}.
These filaments are cross-linked by proteins such as dyneins that undergo a conformational transition through the
hydrolysis of adenosine triphosphate (ATP), so as to slide one filament relative to the other.  Coordinated waves of such sliding under the
constraint that the microtubule bases are held rigid lead to bending waves.  

The anchoring of each axoneme occurs in a special structure called the basal body, which
also serves as a microtubule organizing center during cell division.  Microtubules emerging from each of the two
centers are responsible for pulling the two pairs of chromosomes apart.  Intriguingly, the dual roles for the organizing
centers are mutually exclusive through what is known as the `flagellation constraint', a prohibition against
multitasking: Flagella are lost during cell division.  It has been suggested that the associated loss of motility during
cell division played a role in driving germ-soma differentiation \cite{BellK}.

As there are several excellent historical and recent reviews on the swimming
of microorganisms \cite{Brennen_ARFM,LaugaPowers,StockerARFM}, the present article focuses 
instead on a series of case studies in biological fluid dynamics from
the past decade in which the volvocine green algae have been key to substantial progress.  These include the first direct measurements
of the flow fields around freely swimming microorganisms, studies of nutrient uptake at high P{\'e}clet numbers, findings on
the nature of interactions of swimmers with surfaces, quantitative studies of how eukaryotic flagella synchronize, elucidation of
the mechanism of phototaxis in multicellular swimmers, and exploration of the statistical properties of passive tracers in
suspensions of motile microorganisms. 

I close this section by pointing out that another set of organisms, the choanoflagellates, is emerging as
a model for understanding the transition to multicellularity \cite{King2004}.  Choanoflagellates are understood to be the closest living 
unicellular ancestor of current metazoans (animals) and can exist in both unicellular and multicellular forms.  
Their fluid dynamics is just beginning to be studied quantitatively \cite{Pettitt_choano,Blake_choano,RoperPRL}.

\section{FLOW FIELDS AROUND SWIMMING MICROORGANISMS}
\label{flows}

Our understanding of the swimming dynamics of microorganisms has advanced over the past few decades in part
through the development of new technologies.  Berg's (1971) invention of the tracking microscope  
enabled the discovery of how the run-and-tumble locomotion of peritrichously flagellated bacteria such as {\it Escherichia coli} 
is coordinated to achieve chemotaxis \cite{BergBrown}.
Rather than visualizing organisms through a low-magnification objective with a wide field of view, and hence with
low spatial resolution, this device continuously re-centers the organism in a small field of view at high magnification,
via feedback control of a motorized stage.  Likewise, the advent of relatively inexpensive, high-sensitivity and high-speed video cameras
and specialized fluorescent dyes
has made it possible to study the rapid dynamics of flagellar bundling and unbundling at high temporal resolution, 
uncovering conformational transitions between different helical structures \cite{TurnerRyuBerg}.  These and many other
innovations have made it possible to quantify the stochastic {\it trajectories} of bacteria and other swimming microorganisms.

Only recently, however, has attention turned to actually measuring the flow fields around individual organisms, using
the method of particle image velocimetry (PIV) \cite{PIV}. 
Although such measurements have been made for larger organisms, including fish \cite{Muller} and copepods \cite{Catton}, 
extending the use of PIV down to the micrometer scale has
a number of difficulties.  First, there is the intrinsically stochastic nature of most microorganism swimming so that, even if the organism
swims continuously within the focal plane, it is constantly changing direction.  For example, in the run-and-tumble locomotion 
of {\it E. coli}, the cell swims in an approximately straight trajectory, with its multiple helical flagella bundled behind it, but
changes direction after approximately $1$ s through a tumble, when one or more motors reverse direction and the bundle flies apart. 
If we wish to use the motion of advected microspheres to trace the fluid flow, then it is necessary to change the local frame
of reference to account for such turns.  This can be dealt with (see below) but requires careful data analysis. 
Second, the organisms rarely stay
within the microscope focal plane for long, particularly if undergoing run-and-tumble locomotion, thus requiring the acquisition
of multitudes of very short video clips.  To avoid this problem one can confine the swimmers to
quasi-two-dimensional chambers, but wall effects can then become significant and difficult to incorporate into the analysis. 

It is in this context that one can first appreciate the role of the green algae in biological fluid dynamics.   
Let us consider first the
case of {\it Volvox}.   The earliest measurements of flow fields around {\it Volvox} appear to be those of \textcite{HandHaupt}, 
who were interested in the mechanism of phototactic response to light (discussed further in Section \ref{phototaxis}).  
Their manual tracking of the motion of suspended microspheres near spheroids yielded the first estimates of flow speeds near the
colony surface when it is held in 
place on a microscope slide by a coverslip.
To go beyond these crude measurements, researchers were motivated by the use of micropipettes to hold and study 
giant unilamellar vesicles \cite{Evans}, quasi-spherical lipid bilayer
vesicles that can be tens of micrometers in diameter, and by the world of {\it in vitro} fertilization, in which oocytes are immobilized
with holding pipettes that have been specially shaped to avoid damage to the specimens.  It proves straightforward to hold {\it Volvox} colonies 
on micropipettes with gentle suction, or indeed to exert greater suction to investigate the mechanics of distortion of the ECM 
(see Fig. \ref{fig2}).  
Seeding the surrounding fluid with microspheres enabled a quantitative measurement of the
velocity field as a function of position around the colony surface and its variation with colony size \cite{Multicellular,Flagflows}.
The observed flows can exceed $500$ $\mu$m/s, establishing that these organisms live in the regime of
large P{\'e}clet numbers (Section \ref{life}).    Perhaps even more importantly, the observed dependence 
of the tangential fluid velocity field on the angular coordinate
from the anterior to posterior pole was highly reminiscent of that in
the celebrated squirmer model of \textcite{Lighthill}.   The existence of this biological realization of squirmers 
has made it possible to test many aspects of theoretical predictions arising from that model.

\begin{figure*}[t]
    \begin{center}
        \includegraphics*[clip=true, width = 1.95\columnwidth]{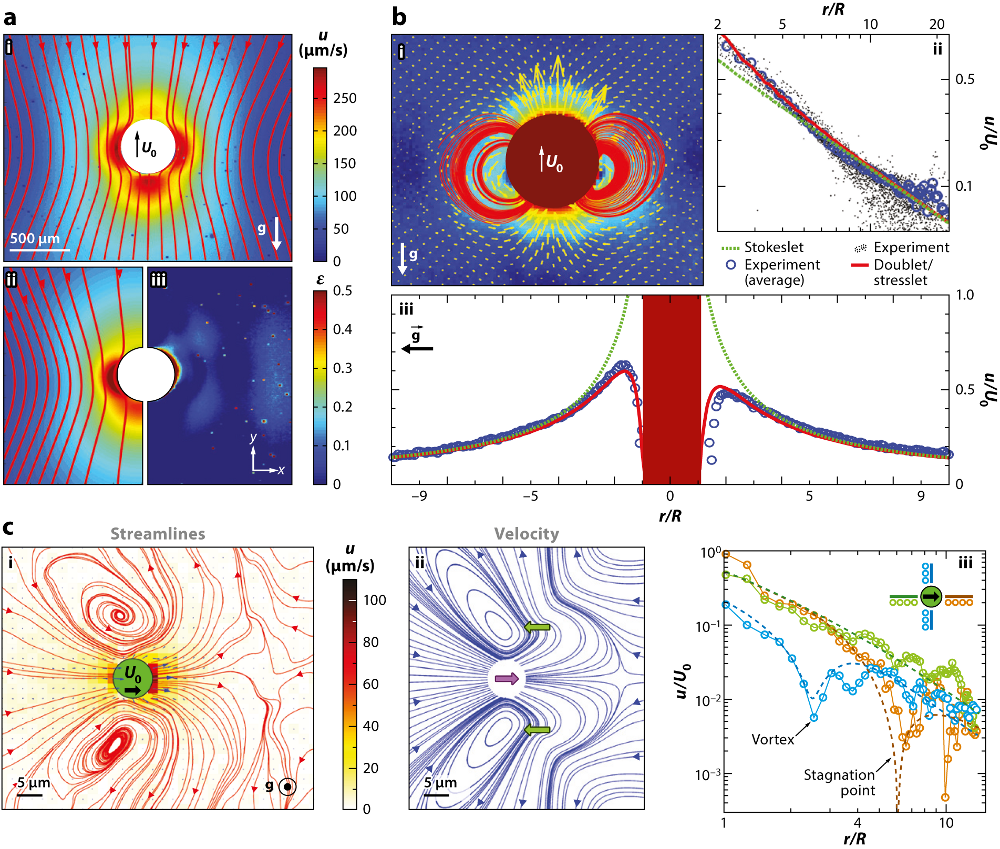}
    \end{center}
    \caption{Flows around freely swimming green algae. ({\it a}) Velocity field around {\it Volvox carteri} in the laboratory frame. ({\it i,ii}) Magnitude and 
streamlines of {\bf u} and theoretical fit, with a common color bar. ({\it iii}) The relative error of fit. The direction of gravity is indicated. ({\it b}) 
Near-field flow around 
{\it V. carteri}. ({\it i}) Magnitude, vector fields, and streamlines of {\bf u} after subtracting a fitted Stokeslet, with the color bar as in part {\it i} of panel {\it a}. 
({\it ii}) Velocity magnitude taken along a horizontal cross section through the center of a colony. An average Stokeslet ({\it green dashed line}) follows the 
observed decay ({\it blue circles}) averaged over the sample set of colonies ({\it black dots}). Deviations from a pure Stokeslet appear at distances less than 
$\sim 5R$ and are accounted for by adding a source doublet and a stresslet ({\it red solid line}). ({\it iii}) Vertical section of {\bf u} through the center of a 
colony, with symbols as in panel {\it ii}, showing front-back asymmetry due to the stresslet. ({\it c}) Average flow field of {\it Chlamydomonas reinhardtii}. 
({\it i}) Streamlines ({\it red}) computed from velocity vectors ({\it blue}). Color scheme indicates flow speed magnitudes. ({\it ii}) Streamlines of the 
azimuthally averaged three-Stokeslet model. Flagellar thrust results from two Stokeslets ({\it lateral green arrows}), whose sum balances the drag 
on the cell body ({\it central purple arrow}). ({\it iii}) Decay of the velocity magnitude for the three directions indicated by separate colors in the inset 
compared to results from the three-Stokeslet model ({\it dashed lines}). Figure adapted with permission from Drescher et al. (2010a). Copyright 
2010 by the American Physical Society.}
    \label{fig4}
\end{figure*}

To go beyond measurements of the flows around immobilized colonies, one can take advantage of the bottom heaviness of  
{\it Volvox}, which results from the positioning of the daughter colonies in the posterior hemisphere of the spheroid 
(which is at the upper right of the colony shown in Fig. 2h).
Such an asymmetrical distribution of mass is responsible for the phenomenon of gyrotaxis, the
combination of swimming and reorientation by fluid vorticity \cite{PedleyKesslerARFM}.  In the absence of any external
fluid flow {\it Volvox} will right itself like a ship with a heavy keel and swim upward against gravity.   A tracking microscope can
then be built that uses a motorized microscope stage turned vertically to control the position of a lightweight video camera that
visualizes an organism as it swims upward in a fixed chamber through fluorescent microspheres illuminated 
by a laser light sheet.  
This enables PIV {\it in the frame
of reference of the swimmer}. The steadiness of {\it Volvox} swimming has allowed for averaging over up to $50,000$ video
frames to get extremely accurate flow fields \cite{Direct} (Figure \ref{fig4}).

For many swimming organisms (bacteria, spermatazoa, uniciellular algae), it is convenient to classify their structure based on the
far-field behavior of the flow they produce, which is that of a stresslet [i.e. two force monopoles (Stokeslets) separated by a
small distance].  Configurations with flagella behind the body are known as pushers, whereas those with flagella in front, 
such as {\it Chlamydomonas}, are pullers.  {\it Volvox} has such a high degree of symmetry that it does not have an 
appreciable stresslet component; instead, the
far field is dominated by the Stokeslet contribution owing to its significant density relative to the surrounding fluid.  
Higher-order singularities become important
close to the colony surface.  One of the key findings in the case of {\it Volvox} was that the Stokeslet is the dominant force 
singularity observed beyond approximately one body diameter.  This finding turns out to be crucial in understanding the wall-mediated effects 
described in Section \ref{surface} and in the problems of tracer statistics in suspensions discussed in Section \ref{tracers}.
Recent work has explored the decomposition of these flows into irreducible representations \cite{Ronojoy}.

In extending such PIV measurements down to the unicellular level, {\it Chlamydomonas} has many advantages over
bacteria.  First, with a body size of $10$ $\mu$m in diameter, it is straightforward to track at moderate magnification, thus
eliminating the need for a true tracking microscope.  Second, we know from three-dimensional 
tracking studies \cite{PolinScience} (see Section \ref{synchro}) that its two flagella, each $10-12$ $\mu$m in length
and beating at $\sim 60$ Hz in a characteristic breaststroke waveform, achieve synchrony persisting for a characteristic time
on the order of $10$ s.  Measurements of the mean-square displacement of free-swimming cells show a crossover from ballistic 
behavior below this timescale to diffusive behavior beyond \cite{Rafai_viscosity}. This long persistence time of straight 
swimming enormously increases the 
likelihood of in-focal-plane trajectories.   Finally, cell bodies can be tracked via their natural
autofluorescence, along with fluorescent microspheres used for PIV.  
Analysis of very large amounts of data yields the time-averaged velocity field shown in part i of Figure \ref{fig4}c.   
We can recall that {\it Chlamydomonas}
spins around its axis at $\sim 2$ Hz \cite{Sourcebook}, so the time average is over not only the flagellar waveform but 
also the orientation of the cell.
There are several important features of this flow field.  First, it is similar to that of a puller stresslet far away from the cell, pulling fluid in from
the front and back, sending it out along the sides of the cell.  But closer to the cell body there is more structure, including the
swirls from the two flagella and a stagnation point in front of the cell.  If we associate a point force with the cell body and
one each at the midpoint of each flagellum, we arrive at the simplest singularity model of {\it Chlamydomonas}, whose
flow field (part ii of Fig. \ref{fig4}c) is remarkably close to the measured one and exhibits the observed spatial decay 
(part iii of Figure \ref{fig4}c).  This provides 
important evidence in support of point force models
for flagella, particularly in the context of synchronization (Section \ref{synchro}).
In independent and contemporaneous work, \textcite{GuastoGollub_timedep} succeeded in measuring the time-resolved
velocity fields around {\it Chlamydomonas} cells swimming in a thin free-standing fluid film.    Using the standard result for the rate of
power dissipated in a two-dimensional fluid of viscosity $\mu$, $P=\int dA 2\mu {\bf E : E}$, 
where ${\bf E}=(1/2)[{\mathbf \nabla u} + {\mathbf \nabla u}^T]$ 
is the rate of the strain tensor, they found peak powers of approximately $15$ fW and average powers through the cycle of 
approximately $5$ fW.  This is consistent with $P=Fv$, where $F$ is the Stokes drag force $F=6\pi\mu R v\sim 10$ pN on a sphere of radius 
$5$ $\mu$m moving at a speed $v \sim 100$ $\mu$m/s, provided one accounts for the average of the dissipated power
being approximately four times the dissipation from the average flow. These results are consistent with estimates of forces and powers 
based on analysis of the actual flagellar waveforms \cite{Antiphase}.  These numbers give interesting insight into the
consumption of ATP by beating flagella.  Using the peak power, the duration of the power stroke, and the hydrolysis energy
per ATP molecule, one finds that the consumption of ATP molecules is on the order of $5,000$ per axoneme per beat, or 
approximately $1$ every few nanometers along the axoneme.  Given the microstructure of the axoneme this is far less than full capacity.

Flow field measurements were ultimately extended to the case of bacteria \cite{Scattering}, using a smooth-swimming 
mutant of {\it E. coli} (HCB437) that does not tumble.  This removes one of the key impediments to tracking but still requires 
the collection of many short-duration video clips.  The resulting flow field is that of a pusher stresslet, with clear 
evidence of the $r^{-2}$ decay of the flow field, and an
amplitude of the stresslet characterized by a 
Stokeslet strength of $0.42$ pN and a separation of $1.9$ $\mu$m, very consistent with expectations for an organism the size of
{\it E. coli}.  By restricting measurements to trajectories close to a no-slip surface, one can also see the modifications to the flow field 
due to that
boundary, which are fully consistent with calculations of stresslets near surfaces \cite{Blake}, most importantly
demonstrating the weaker $r^{-3}$ decay and multilobed structure.

\section{LIFE AT HIGH P{\'E}CLET NUMBERS}
\label{life}

All the organisms considered in this review live in a world of small or even ultrasmall Reynolds numbers, $Re=UL/\nu$.
Even for the largest Volvocales, with $L\sim 0.1$ cm and a swimming speed of at most $U\sim 0.1$ cm/s, we have $Re\sim 1$.  
But the neglect of advective contributions to momentum transport does not carry over to the transport of molecular species.
In this case, the relative importance of advection to diffusion is given by the P{\'e}clet number, $Pe=UL/D$.
Because the diffusivity $D$ of small molecular species is typically $\sim 10^{-5}$ cm$^2$/s, three orders of magnitude smaller than
the kinematic viscosity of water, the P{\'e}clet numbers are that much larger.  Thus, even for {\it Chlamydomonas} we
find $Pe\sim 0.1-0.5$ \cite{TamHosoi}, whereas for {\it Volvox}, it can be several hundreds \cite{Multicellular,Flagflows}.   
This is a very different regime than that for the case of bacteria, with $L\sim 1-5$ $\mu$m and $U\sim 10-40$ $\mu$m/s, so 
$Pe\sim 0.01-0.2$, as emphasized by \textcite{BergPurcell} in their classical work on
chemoreception.

One of the most basic possible implications for organisms living in a world in which advection matters is a change in
the rate of nutrient uptake or waste dispersal, relative to the purely diffusive behavior.  
The first experimental evidence that fluid flow would matter for the metabolism of {\it Volvox} colonies came from comparative 
studies of germ cell growth under normal conditions \cite{Multicellular}, deflagellation of the mother colony, 
the removal of germ cells in 
a quiescent fluid,
and one in which artificial stirring was provided.  It was shown that a lack of external fluid flow reduced germ cell
growth, which could be rescued by stirring.
Later work showed significantly stronger evidence of phenotypic plasticity in the flagellar apparatus 
in the presence of nutrient deprivation for {\it Volvox} than for {\it Chlamydomonas} \cite{Plasticity}.

Our discussion of theoretical approaches to nutrient uptake at large $Pe$ begins by noting that the transport of a scalar quantity 
in the presence of flow is a classic problem in fluid mechanics, first studied in detail, with heat conduction providing
a physical and mathematical analog for diffusion, using the
method of matched asymptotics in the context
of heat transport from a solid sphere in a laminar flow \cite{Acrivos_Taylor}.
In a quiescent system, neglecting buoyancy effects, the current of heat $Q$ is given by the diffusive flux, itself calculated
from the gradient of the equilibrium temperature field deviation $T(r)=T_0(1-R/r)$, so $Q\sim R$.
At very large $Pe$, a boundary layer develops in which there are enhanced thermal gradients at the front of the sphere and a
long trailing thermal wake.  A balance between the local advective radial flow and diffusion leads to a 
boundary layer thickness $\delta\sim Pe^{-1/3}$ and the flux enhancement $Q\sim R Pe^{1/3}$.   

Organisms such as {\it Volvox} have substantial fluid flow over their surfaces driven by flagella, and although
there is a no-slip condition, one expects the uptake rate to be different than the non-self-propelled situation.  The first
such calculation was done by Magar and colleagues \cite{Magar1,Magar2} using both steady and unsteady squirmer models, with similar 
results found in a different model with a similar tangential velocity profile \cite{Multicellular}.  The
alteration of the velocity field in the neighborhood of the surface leads to the scaling $\delta\sim Pe^{-1/2}$ and
$Q\sim RPe^{1/2}$.   

A heuristic argument suggests a possible implication for this new scaling law.  An organism with the
architecture of the Volvocales has all or at least the vast majority of its cells on its surface, so its metabolic needs must scale
as the surface area, $\sim R^2$.  Without flagella-driven flows the nutrient uptake rate scales as $R$, implying a bottleneck at
some radius, suggesting that {\it Volvox} could not live by diffusion alone.  Although difficult to estimate, this bottleneck size appears to be in
the middle of the Volvocales size spectrum.  Now, if we adopt a simple model for swimming, in which the total propulsive
force from flagella is proportional to the number of somatic cells, and hence $R^2$, while the drag is proportional to the radius $R$,
we conclude that the swimming speed $U\sim R$, which in turn implies $Pe\sim UR\sim R^2$, and hence $Q\sim R^2$,
removing the diffusive bottleneck.  This result suggests that there is a fitness advantage {\it per somatic cell} to be in a
larger colony that swims faster, providing a potential driving force for increased size, although other factors can surely play a role,
such as the ability to escape from predators.  It also suggests a reason for germ-soma differentiation, for if the cooperative swimming is
needed to enhance nutrient acquisition then sequestering the germ cells inside the colony overcomes the lack of swimming
during cell division (the flagellation constraint).  This idea is complementary to the argument by \cite{BellK} that larger
organisms will sink too far in the water column during cell division, again because of the flagellation constraint, without developing
germ-soma differentiation.  

Looking more generally at the possible surface velocity profiles within squirmer models, 
\citet{Lauga_feeding} recently showed that
optimizing swimming is equivalent to optimizing nutrient uptake for all $Pe$.  Along with the work of \textcite{TamHosoi} on
optimizing waveforms of biflagellated algae for nutrient uptake, we can see the emergence of more quantitative mechanical 
analysis of problems in evolutionary biology.

\section{SURFACE INTERACTIONS}
\label{surface}

In their natural habitats, many swimming microorganisms find themselves close to surfaces, from bacteria in confined 
physiological environments to sperm cells in the fallopian tubes.  On rather general grounds, it would be expected that
nearby surfaces can induce long-range interactions with such swimmers, owing to the slow spatial decay of the low-order 
multipole flow fields of the Stokeslets and stresslets \cite{Blake,BlakeChwang74}.    
The presence of a no-slip wall can also induce large-scale eddy currents that can assist feeding, as first
discussed by \textcite{Higdon_feeding} and elaborated on more recently in several contexts 
\cite{pepperroperstone,Blake_choano}.    This section considers two problems involving the swimming of
organisms near surfaces to illustrate the role the volvocine algae have played in revealing new phenomena and in
testing various theoretical predictions.

\begin{figure*}[t]
    \begin{center}
        \includegraphics*[clip=true, width = 1.95\columnwidth]{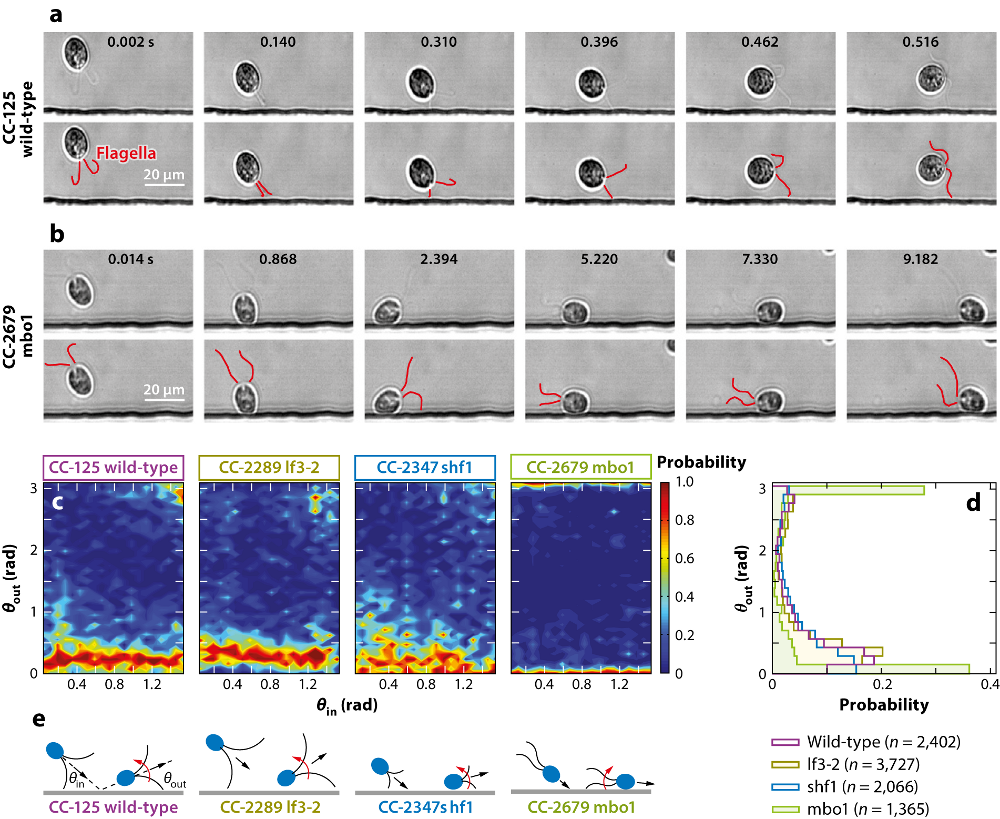}
    \end{center}
    \caption{({\it a}) Scattering process for wild-type {\it Chlamydomonas} CC-125: (top row) original images and (bottom row) flagella highlighted in red. 
Results for the long-flagella mutant lf3-2 and the short-flagella mutant shf1 appear qualitatively similar. The scale bar is 20 $\mu$m. ({\it b}) As in panel 
{\it a} but for the mutant pusher mbo1 (which moves backward only), which remains trapped for several seconds. ({\it c}) The conditional probability 
distributions $P(\theta_{out}\vert\theta_{in})$ indicating that, for all four strains, the memory of the incident angle is lost during the inelastic collision 
process owing to multiple flagellar 
contacts with the surface. ({\it d}) The cumulative scattering distribution  $P(\theta_{out})$ showing how cilia length and swimming mechanisms determine 
the effective surface-scattering law. ({\it e}) Schematic illustration of the flagella-induced scattering and trapping mechanisms. Figure adapted with 
permission from Kantsler et al. (2013).}
    \label{fig5}
\end{figure*}

\begin{figure*}[t]
    \begin{center}
        \includegraphics*[clip=true, width = 1.95\columnwidth]{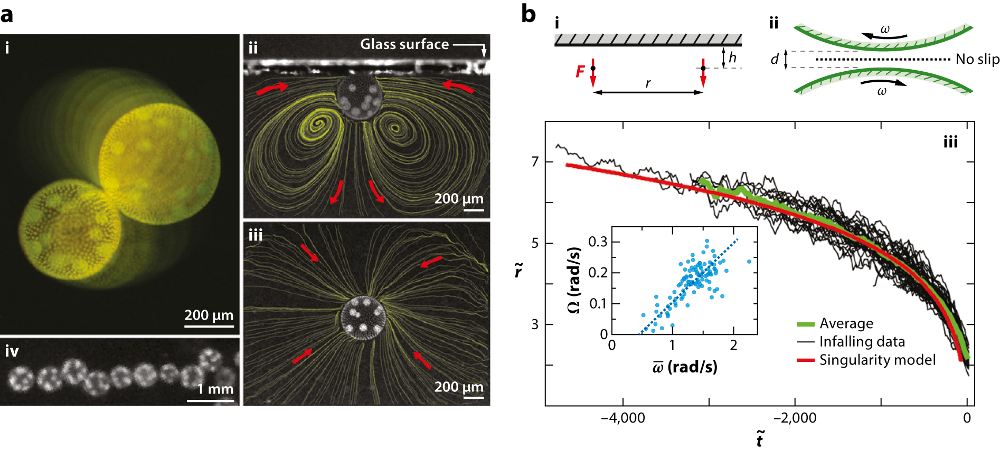}
    \end{center}
    \caption{({\it a}) Observations of the hydrodynamic bound state of {\it Volvox}. ({\it i}) View through the top of a glass chamber showing superimposed 
images taken 4 s apart, graded in intensity for clarity. ({\it ii}) Side and ({\it iii}) top views of a colony swimming against a coverslip, with fluid streamlines 
from particle image velocimetry. ({\it iv}) A linear cluster viewed from above. ({\it b}) Data and analysis of the hydrodynamic bound state of {\it Volvox}. 
Geometry of ({\it i}) two interacting Stokeslets (side view) against an upper no-slip surface and ({\it ii}) the gap between nearby spinning colonies. 
({\it iii}) Radial separation {\it r} of infalling colonies, normalized by the mean colony radius, as a function of rescaled time for $60$ events ({\it black}). 
The running average ({\it green}) compares well with predictions of the singularity model ({\it red}). ({\it Inset}) Orbiting frequency $\Omega$ versus 
mean spinning frequency $\bar{\omega}$ , and the linear fit. Figure adapted with permission from Drescher et al. (2009b). Copyright 2009 by the 
American Physical Society.}
    \label{fig6}
\end{figure*}

In 1963, Rothschild observed that sperm cells accumulate near the walls of a container, an effect seen more recently
with bacteria \cite{Berke}. 
There have been two contrasting explanations for this effect: one based on long-range hydrodynamic interactions 
\cite{Berke} and another based on contact interactions with the wall \cite{Tang}.   The first argument starts
from the picture of sperm or bacteria as generating far-field flows associated with pusher stresslets, and examines the
effects of a no-slip wall on both the orientation and distance from the wall.  A straightforward calculation shows that
cells swimming parallel to the wall experience an attractive interaction decaying with distance as $z^{-3}$ and torques
that will tend to turn non-parallel orientations toward parallelism, at a given separation.   Assuming that the parallel 
orientations are achieved, it is possible to balance the attraction against diffusion to arrive at a steady-state profile
that shows accumulation near walls, as in experiments, although the quantitative agreement at short distances is only fair.

The second approach argues that long-range hydrodynamics play little role: instead, the direct contact
interaction with the surfaces is crucial.  Here the picture is one in which the incoming and
outgoing angles
of a swimmer approaching and then leaving a wall are very different.    If most incoming orientations lead to shallow outgoing
angles, then cells will tend to accumulate near the walls, with rotational diffusion
randomizing the outgoing trajectories.

Understanding when the effects of long-range hydrodynamic interactions on cellular reorientation dominate over
rotational Brownian motion is an important general issue not only in cell-surface interactions, but also in cell-cell 
scattering.   A simple argument suggests that there is a characteristic length scale, termed the
hydrodynamic horizon, beyond which hydrodynamic effects are negligible relative to rotational diffusion \cite{Scattering}.  
Let us consider
a bacterium at the origin, and the far-field velocity field around it,
\begin{equation}\label{e:dipole_model}
{\bf u}({\bf r})
=
\frac{A}{|\mathbf r|^2} \left[3(\mathbf{\hat r}\cdot{\mathbf{d}'})^2-1\right] \mathbf{\hat{r}}
\end{equation}
where $A=\ell F/8\pi \mu$ is the stresslet amplitude, and ${\mathbf{d}'}$ is the unit vector in the swimming direction.
The position and orientation of a second swimmer, at position ${\mathbf{x}}$ and 
moving with speed $V_0$ in direction ${\mathbf{d}}$, obey
\begin{eqnarray}
\mathbf{\dot{x}}
&=&
V_0 \mathbf{{d}}  +  \mathbf{u},
\label{e:wall_scattering_a}\\
\mathbf{\dot{{d}} }
&=&
\frac{1}{2} \; \mathbf{\omega} \times {\mathbf{d}} +
\Gamma\;
\mathbf{{d}}\cdot \mathbf{E}\cdot (\mathbf{I}-\mathbf{{d}}\mathbf{{d}})
\label{e:wall_scattering_b}.
\end{eqnarray}
where ${\mathbf \omega}$ is the vorticity, and $\Gamma$ is a shape factor for
rod-shaped cells.
If we assume there is an interaction timescale $\tau$ between the two cells,
the angular rotation of the second cell will be
$\Delta\phi\sim \omega \tau$, where  
$\omega$ falls off as $A/r^3$, so  $\langle \Delta \phi(\tau,r)^2 \rangle_H
\sim A^2\tau ^2/r^6$.
The corresponding mean squared angular deviation from rotational Brownian motion will be
$\langle \Delta\phi(t)^2\rangle_D=4D_r \tau$, and a balance between the two effects occurs at the
length scale
\begin{eqnarray}
r_H \simeq \left(\frac{A^2\tau}{D_r}\right)^{1/6}.
\label{eq:rhorizon}
\end{eqnarray}
Estimates for {\it E. coli} \cite{Scattering} suggest that $r_H\simeq 3.3$ $\mu$m, a value that would
require a volume fraction of cells on the order of 
$5-10$\%, which approximates the limiting value at which collective effects are observed.  This line of reasoning suggests therefore that in dilute suspensions and in
the interaction of cells with surfaces long-range hydrodynamic interactions should play a minor role, and more local effects should be dominant.

\citet{Ciliary} tested these ideas by directly visualizing the interactions between swimming microorganisms and surfaces 
in microfluidic chambers.  They used bull spermatazoa as an example of pusher swimmers, and wild-type and mutant strains of 
{\it Chlamydomonas} were the chosen pullers.  The essence of the results is contained in the scattering plots of outgoing versus 
incoming angles (Figure 5), which clearly show the strongly inelastic character of surface interactions.  These can be attributed to 
purely geometric effects, in which the tail undulations for sperm trace out a wedge larger in amplitude than the head diameter, 
thus driving the head into the wall at a slight angle.  This can be modified
by changing the temperature, which alters the beat amplitude and correspondingly changes the probability distribution function.
In the case of {\it Chlamydomonas} the high-speed videos show directly that cells interact with the wall through flagellar contact and depart from the wall at
a shallow angle determined by the geometry of the flagellar waveform relative to the body size (Figure \ref{fig5}).  
Experiments with mutants having shorter flagella and those
moving only backward are consistent with this geometric picture.


\citet{Squires} pointed out one of the more intriguing consequences of the presence of a no-slip boundary on a nearby Stokeslet \cite{Blake}.
Building on earlier work  on hydrodynamically induced attractive interactions 
between colloidal particles near walls \cite{like_charge}, he
observed that if it were possible to levitate a pair of spheres near the upper surface of a container, so that their
individual Stokeslets (due to gravity) pointed {\it away} from the ceiling, there would be a weak \textit{attractive} interaction 
between the two, mediated by the no-slip surface.  Although he imagined something like an electrophoretic mechanism to achieve 
the levitation of passive particles, {\it Volvox} naturally swims upward because of its bottom heaviness and thus shares this
behavior.
Using a tracking system designed for protists, \textcite{Tracking} stumbled upon a realization of Squires' calculation while
investigating the possibility of complex pairwise interactions between spherical swimmers, motivated by simulations on
suspensions of squirmers \cite{IshikawaPedley}.

As an aside, I note that the tracking system used for these experiments addresses some of the key difficulties encountered when studying the swimming of large 
protists.  If one wishes to examine swimming far from walls, then the experimental chamber must be many times the organism's 
diameter, approaching at least
the centimeter scale.  To avoid triggering a phototactic response by the organisms, one must use red light illumination, 
but this, combined with the
larger chamber size, can produce thermal convection currents that overwhelm the swimming cells.  Great care 
must be taken to avoid this by suitable stirring of an external heat bath.  

Drescher et al.'s discovery was termed a hydrodynamic bound state, and its basic properties are illustrated in Figure \ref{fig6}.  
Two nearby {\it Volvox} swimming up to the top surface of the chamber are attracted together and orbit each other in a stable state.   
Although the orbiting is a particular feature of
{\it Volvox}, the infalling of the two colonies toward each other is in precise agreement with theory, provided one treats the interactions between
the two colonies as those of two Stokeslets. The measurements of the flows around freely swimming {\it Volvox} discussed in Section \ref{flows} 
provide justification for this approximation.
When the two {\it Volvox} are up against the chamber top, the only degrees of freedom are their lateral positions ${\bf x}_i$, which 
evolve according to
$\dot{\bf x}_i={\bf u}({\bf x}_i)$, where ${\bf u}({\bf x}_i)$ is the local velocity at colony $i$ due to the second colony.  
The upswimming speed $U$ of colonies
can be readily measured, as can their settling speed $V$ when deflagellated, so the Stokeslet strength is simply $F=6\pi\mu R (U+V)$, 
and the distance $h$ of
each Stokeslet from 
the wall is the colony radius $R$.   If we rescale the radial distance between colonies as $\tilde r=r/h$ and time as $\tilde t=tF/\mu h^2$, then the infalling
trajectories obey
\begin{equation}
\frac{d\tilde{r}}{d{\tilde t}}=-\frac{3}{\pi}\frac{{\tilde r}}{(\tilde{r}^2+4)^{5/2}}~.
\end{equation}
The excellent agreement with the data shown in Figure \ref{fig6} is strong evidence for this wall-mediated flow mechanism.  
The analysis has been extended to account for
the orbiting dynamics of the colonies by an application of lubrication theory, and a second type of bound state that occurs when 
colonies hover near the chamber 
bottom.  Sophisticated conformal
mapping methods have recently been developed to treat more complex examples of hydrodynamic bound states \cite{Crowdy}.

\section{SYNCHRONIZATION OF FLAGELLA}
\label{synchro}

One of the long-standing issues in the swimming of microorganisms is the manner in which flagella synchronize.
A key historical point is the observation by \textcite{Gray}, \textcite{Rothschild}, and others that nearby swimming sperm 
cells tend to synchronize their
beating, and this was the motivation for Taylor's (1951) celebrated model of waving sheets discussed
below, although 
synchronization occurs in both prokaryotic and eukaryotic systems \cite{Dance}.  In the former case, the helical 
flagella of peritrichously
flagellated bacteria bundle and unbundle as the rotary motors that turn them stochastically switch direction.  In the
unbundled state, a bacterium executes a tumble, whereas when bundled, the cell body moves forward in a straight line.
This is the run-and-tumble locomotion that has been much studied.  Although the underlying biochemistry of power
generation and switching regulation is known in exquisite detail for bacteria, the bundling dynamics {\it in vivo} are not 
[although there are elegant macroscopic studies of this problem with mechanical models \cite{Powers_macro}].
The eukaryotic case is  the precise opposite in many ways, for there is now emerging a detailed body of work on
the synchronization dynamics, but the (admittedly more complex) biochemistry of beat generation and regulation is still unclear.
We focus on the eukaryotic situation here.

No discussion of synchronization is complete without reference to the famous work of \textcite{Huygens} 
on the synchronization of pendulum clocks, reproduced
in a modern setting by \textcite{Schatz}.  Huygens discovered that nearby clocks supported on a common structure would 
phase lock, typically in 
antiphase, owing to vibrations in the structure.  In modern language we would say that the two
oscillators synchronized through a weak, presumably linear coupling.  The fundamental question in the
biological context can be stated as, What is the biological equivalent of the
wall coupling Huygens's clocks?  One of the first key insights was provided by \textcite{Taylor_sheets}, whose work
sparked the search for a hydrodynamic mechanism.

Taylor's work on waving sheets abstracted the problem of synchronization to the analysis of the fluid dynamics of two parallel,
laterally-infinite sheets of an inextensible material subjected to sinusoidal traveling waves of vertical displacement.  
By calculating the rate of energy dissipation in the surrounding fluid as a function of the relative phase of the two 
waves he found that there is a minimum in the dissipation when the waves were nested (i.e., in phase).   
This was not a dynamical 
calculation by which synchronization was obtained from an arbitrary initial configuration, although later work has done this \cite{Fauci}, 
including that on the effects of sheet flexibility \cite{Lauga_synchro}.
Of course, the minimization of dissipation is not a general physical principle from
which to derive behavior; it may be true in a given situation, but it is not a satisfactory approach in general.  We expect 
that one needs to consider the response of the axoneme as a whole (motors, filaments) to forces and torques in order to understand synchronization.  
It is important to emphasize the distinction between the mechanism of the coupling (hydrodynamic, elastic) and the
mechanism of response of the flagella to the forces and torques produced by a neighboring filament.  The latter could
be related to the molecular motors themselves, to mechanosensitive ion channels in the flagellum, or other
as yet unknown biological features.

\subsection{Pairs} 

The most important early work on the synchronization of eukaryotic flagella was done by 
\textcite{RufferNultsch1,RufferNultsch2,RufferNultsch_ptx1,RufferNultsch_ptx1a}.
Using high-speed imaging techniques of the day (images captured on film) and manual light-table tracings of individual
movie frames, they established the essential features of
flagellar synchronization for both wild-type and a mutant of {\it Chlamydomonas}.
Although their methods lacked connection with the more recently developed viewpoint of stochastic nonlinear oscillator theory,
and their observations were not fully quantitative, their essential findings are seminal.

\begin{figure*}[ht]
    \begin{center}
        \includegraphics*[clip=true, width = 1.4\columnwidth]{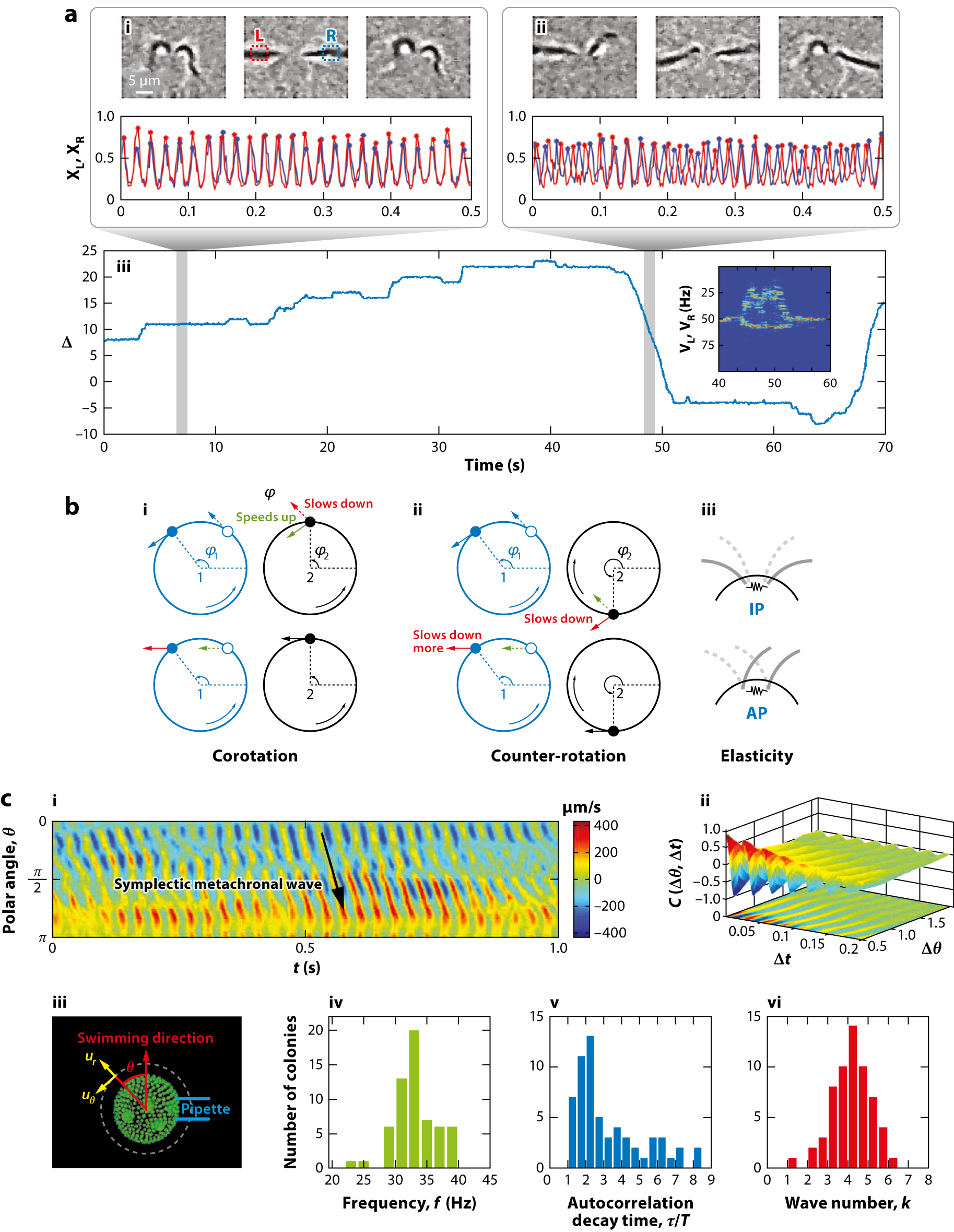}
    \end{center}
    \caption{({\it a}) Synchronization in {\it Chlamydomonas}. ({\it i,ii, top}) Images from a high-speed video of flagellar beating in synchrony and 
during asynchrony, with interrogation windows for Poincar{\'e} sections in part {\it i}. ({\it i, ii, bottom}) Time series of left and right pixel intensities 
in windows. ({\it iii}) Time series of phase difference $\Delta$ over $70$ s, showing periods of synchrony, slips, and drifts. ({\it Inset}) Fourier transform 
during transition from synchrony into drift and back. Panel {\it a} adapted with permission from Polin et al. (2009). ({\it b}) Mechanisms of synchronization. 
In the elastohydrodynamic mechanism, ({\it i}) corotation leads to in-phase (IP) synchrony, and ({\it ii}) counter-rotation produces antiphase 
(AP) locking. ({\it iii}) Elastic couplings can provide a competing effect. Panel {\it b} adapted with permission from Leptos et al. (2013). 
({\it c}) Metachronal waves in {\it Volvox carteri}. ({\it i}) Space-time illustration of a radial component of the fluid flow measured near the colony 
surface.  ({\it ii}) Correlation function of waves. ({\it iii}) Confocal microscope image of {\it Volvox}, showing the colonial axis and the radius for 
velocity components (dashed). Histograms of ({\it iv}) beat frequencies, ({\it v}) decay times of the correlation function scaled by the beat period, 
and ({\it vi}) the metachronal wave number for $60$ colonies. Panel {\it c} adapted with permission from Brumley et al. (2012).}
    \label{fig7}
\end{figure*}

One of their most important findings was the existence of 
three distinct modalities of beating; synchrony, synchrony with
transient interruptions, and asynchrony \cite{RufferNultsch2}.  The synchronous state (seen $\sim\!85\%$ of the time) is the familiar 
breaststroke, in which
there may be slight variations in the period from beat to beat, but the two flagella appear phase locked.  The
interrupted synchrony ($\sim\!10\%$) took the form of several faster  beats of one flagellum ({\it trans}; see
Figure \ref{fig3}), which would then 
re-synchronize with the {\it cis} flagellum.  In the asynchronous state ($\sim\!5\%$), the two flagella had different
frequencies and no phase locking.    These observations were made with individual movies lasting
only a few seconds, and the authors proposed that these different beating patterns corresponded to three distinct
subpopulations of cells.

One of the features that makes {\it Chlamydomonas} so useful in the study of eukaryotic flagella is the relative ease with
which it can be studied by micropipette manipulation, the method of choice for R{\"u}ffer \& Nultsch.  With
digital high-speed video microscopy it is now possible to obtain long time series of flagellar beating and to quantify 
synchrony in a way that allows quantitative comparison with theory \cite{PolinScience}.   
Of the many ways to quantify the beating dynamics,
the simplest and most straightforward is essentially a Poincar{\'e} section.   Small interrogation windows are created within
the digital images on either side of the cell body, and the pixel intensity in each is examined over time (Fig. \ref{fig7}).  A criterion based
on those intensities can be used to define the discrete times at which the phase angle $\theta_i$ of 
each flagellar oscillator reaches a
multiple of $2\pi$, with linear interpolation of the phases in between those times.  From these, one can construct the
normalized phase difference $\Delta=(\theta_1-\theta_2)/2\pi$.  Periods of synchrony correspond to plateaus in
$\Delta(t)$, steps up or down by integers correspond to the transient asynchronies (henceforth called
slips) that R{\"uffer} \& Nultsch found, and asynchronous beating is marked by $\Delta(t)$ locally linear in time (termed drifts).   
Part iii of Figure \ref{fig7}a
shows a typical trace of $\Delta(t)$ over a bit more than $1$ min, and we see all three behaviors R{\"u}ffer \& Nultsch found but in a
single cell!  Averaging over many cells shows the fraction of time spent in the three modes of beating to be
$0.85:0.10:0.05$ within experimental error.   This is a state of affairs just like in statistical physics, where ensemble
averages (over many cells) and time averages (on a single cell) yield the same results.  

In Figure \ref{fig7}, we can see that the synchronous periods are noisy.  One can easily calculate the expected thermal fluctuations
of a passive axoneme based on its length and stiffness and deduce that the observed noise level is much larger.
This is biochemical noise associated with the underlying action of molecular motors distributed along the length
of the axoneme.  
Whether one has a reduced description of flagellar dynamics in terms of phase angles or complete waveforms, a 
fundamental question is how to extract useful information about the underlying biomechanics of
beating and synchronization  from time series.  We discuss here one analysis of phase-angle dynamics that has proven
useful.  In the absence of any coupling between the oscillators, their phase angles would evolve as
$\dot\theta_i=\omega_i$, so $\dot\Delta=\delta\nu$, where 
$\delta\nu$ is the intrinsic frequency difference $(\omega_1-\omega_2)/2\pi$.  Any coupling between the two
oscillators must respect the periodicity of the angles and, in the simplest case, would depend only on the angle
difference $\Delta$.  As seen above, the actual time series of $\Delta$ is noisy, and the equation of motion
for $\Delta$ should be a stochastic ordinary differential equation (ODE), the simplest case being additive noise on top of a deterministic 
dynamics.  An example of the latter is the famous \textcite{Adler} equation, first derived
in the context of the synchronization of electronic oscillators.  Introducing a random noise $\xi(t)$, we obtain
\begin{equation}
\dot\Delta=\delta\nu-\epsilon\sin(2\pi\Delta)+\xi(t)~,
\end{equation}
where $\epsilon$ is a coupling parameter for the lowest-order periodic term.   
We take $\langle \xi\rangle=0$ and $\langle \xi(t)\xi(t')\rangle=2T_{\rm eff}\delta(t-t')$, where angular
brackets denote an average over realizations of the noise and $T_{\rm eff}$ can be thought of as an 
effective temperature.

To estimate $\delta\nu$, $\epsilon$, and $T_{\rm eff}$ from the time series, one notes
that the stochastic Adler equation also reflects the motion of a Brownian particle on a tilted washboard potential.
Periods of synchrony correspond to trajectories near one of the minima, slips correspond to thermally assisted hopping from
one minimum to an adjacent one, and drifts are found in the regime in which the tilt is sufficiently large that there
are no longer any minima, so the particle runs downhill.  If the noise level is not too large, then the fluctuations near
any given minimum occur in a potential well that is well approximated by a quadratic, and one can appeal to known
results to obtain three independent observables from which to determine the unknowns.  These are the amplitude
and decay time of the autocorrelation function of $\Delta$,
\begin{equation}
R_0 = \tau_{\mbox{{\scriptsize ac}}} \,T_{\mbox{{\scriptsize eff}}} \quad ; \quad \tau_{\mbox{{\scriptsize ac}}}  
= \frac{1}{2\pi\sqrt{(2\pi \epsilon )^2 -  \delta \nu ^2}}.
\label{autocorrelation_params}
\end{equation}
and the relative left-right hopping probability,
\begin{equation}
p_+/p_- = \exp(\delta \nu/T_{\rm eff}).
\label{jumps_ratio}
\end{equation}
Analysis of large amounts of data yields intrinsic frequency differences that are smaller than a few percent in
the synchronized regime but are as large as $20\%$ during drifts.  We conclude that {\it Chlamydomonas} is able to
change the internal frequency difference between its two flagella.  When it is small, the interflagellar coupling leads
to synchronous behavior, and when it is large, there are drifts.  A consistency check in this analysis is that the
tilt of the washboard potential implies that there is a small shift in the phase of the locked state from zero, approximately one-eleventh
of a cycle.  This had independently been measured by R{\"u}ffer \& Nultsch.
Subsequent analysis showed that the noisy Adler equation is in quantitative agreement with the
detailed dynamics of a phase slip \cite{NoisePRL}.
 
These observations suggested that individual {\it Chlamydomonas} cells execute a eukaryotic equivalent of
bacterial run-and-tumble locomotion which bacteria do 
by stochastically bundling and unbundling their flagella: run-and-turn locomotion.
If this is the case, then a population should diffuse at a corresponding rate.  Measurements of the diffusion of cells in a suspension yielded
diffusivity $D\sim 10^{-3}$ cm$^2$/s.  For a stochastic process of the type imagined, we expect $D\sim u^2\tau$, where
$u$ is a typical swimming speed and $\tau$ is the mean time between turns.  For {\it Chlamydomonas}, we
know that $u\sim 100$ $\mu$m/s, so $\tau$ should be $\sim 10$ s.  Measurements of the trajectories of
individual cells indeed showed that there is an exponentially distributed PDF of the time between sharp
turns, with a time constant of $11$ s \cite{Tracking}.    Although models are now emerging to describe this run-and-turn locomotion from a 
microscopic picture of noisy flagellar orbits \cite{BennettGolestanian},
it must be emphasized that at present there is no understanding of the biochemical
origin of this timescale. 

Let us now consider some of the leading proposed mechanisms by which flagella can synchronize, and their
corresponding models.  There are two basic types of mechanisms proposed: (a) waveform compliance and
(b) variable forcing.  The easiest way to visualize these is to first adopt the point of view that the motion of flagella
during a beat cycle can be represented by the motion of microspheres around closed trajectories.  That this can
be a faithful representation of real flagella is seen in the above discussion of the flow fields around
individual {\it Chlamydomonas} cells, for which the three-Stokeslet model provides a surprisingly accurate representation of
the time-averaged flow field \cite{Direct}.   This provides {\it ex post facto} justification for many models of
synchronization.  With regard to waveform compliance, the work of \textcite{Niedermayer} provides a particularly clear derivation of a key
elastohydrodynamic effect.  With the flagella modeled for analytic convenience as spheres driven around circular orbits by a constant tangential force,
deviations from that ideal orbit are allowed by a spring whose equilibrium length is the unperturbed orbital radius.
With a constant driving force, deflections of the sphere to a larger radius orbit lead to a decrease in the angular velocity,
and deflections inward produce higher angular velocities.  It is then straightforward to see how synchronization
occurs when the spheres are coupled through the Stokeslet fields they create (Figure \ref{fig7}b).  The predicted coupling strength then
scales as $\ell^3$, as this is the stiffness of an elastic filament of length $\ell$.  

It is not easy to alter growth conditions to change the equilibrium length of flagella in any systematic way, but nature
has provided us with a powerful mechanism to investigate the dependence of synchrony on the flagellar length.  When flagella
are subject either to a pH shock or to mechanical stress they often are shed and can grow back over 
an hour or two.  This is an eternity on the timescale necessary for synchronization studies, so it is possible to
investigate synchronization at discrete points in time throughout the regrowth phase \cite{Emergence}.  Results show
a very strong dependence on the flagellar length which is consistent with the elastohydrodynamic mechanism.

\textcite{UchidaGolestanian1,UchidaGolestanian2} proposed a second mechanism for synchronization.  It is, in many
ways, the converse of the elastohydrodynamic mechanism, which postulated a constant internal driving force and a
flexible trajectory.  Uchida \&Golestanian studied the case of rigid cyclic trajectories with variable tangential drive along
the path and arrived at general conditions under which synchronization is possible.  From a biological point of view,
there is no doubt that the intraflagellar forces vary considerably during a cycle, for Guasto et al.'s (2010) time-resolved tracking studies
and more recent work \cite{eLife} have quantified the variation between the power
and recovery strokes.    The analysis of synchronization corresponds to that in the elastohydrodynamic picture,
because the Stokeslet flow fields still influence each particle, but their speedup or slowdown happens within
rigid orbits.
Recent work using colloidal particles forced by time-dependent optical traps has tested these two ideas and shown their 
interplay \cite{CicutaPRL1,CicutaPRL2}.  

More recently, it has been suggested that direct hydrodynamic interaction between flagella is not necessary for
synchronization \cite{RockingPRL,RockingPNAS}; instead, the rocking of the cell body when flagella
desynchronize can put them back into synchrony.   This is an interesting idea, although some other mechanism, such as
elastic couplings at the flagellar bases, must be invoked to explain the
synchronization of flagella on pipette-held cells.

As is often the case in biological studies, key insight can be gained from the study of mutants.  In the context of
synchronization, {\it ptx1}, a mutant with deficient phototaxi, is of great interest.  
First isolated by \textcite{Witman_ptx1}, {\it ptx1} is believed to 
lack the asymmetry between the way its two flagella respond to calcium, which is known to be essential to the phototactic response 
(see Section \ref{phototaxis} for
a more detailed discussion).  If its two flagella were nearly identical, {\it ptx1} would be expected to exhibit
distinct behavior from the wild type.  Once again, \textcite{RufferNultsch_ptx1,RufferNultsch_ptx1a} provided 
important early studies, showing that {\it ptx1} displays not only the ordinary breaststroke seen in the wild type, but also
a second mode of so-called `parallel' beating for the motion of the flagellar bases.  This new mode occurs at
a significantly higher frequency than the breaststroke, close to the instantaneous frequency during a phase slip of the wild type.
Using modern high-speed imaging, \textcite{Antiphase} found that this state is precise antiphase synchronization,
with stochastic transitions back and forth between in-phase and antiphase states. Selective deflagellation studies of the wild type show that one 
flagellum by itself can also execute the higher-frequency mode \cite{Lag}.

Some insight can be gained by re-examining the bead-spring model of \textcite{Niedermayer}, but allowing for the
fundamental orbits to be of the opposite direction, as in {\it Chlamydomonas}.  In this case, one finds that the natural synchronized state
is in fact antiphase, whereas it is in phase for the ciliary case (two beads orbiting in the same sense) \cite{Antiphase}.  
This implies that {\it ptx1} is consistent with this
basic elastohydrodynamic model and the wild type is the exceptional case.  Clearly there must be more 
operating in  {\it Chlamydomonas} synchronization than what is captured by these simple models.  Perhaps elastic 
couplings at the flagellar bases are important, possibly involving the striated filaments that
run between the basal bodies \cite{Lechtreck}. It should also be noted that the plausible assumption that an 
increase in radius produces a slowdown in angular
velocity need not hold.  If the reverse is true then it is possible to find antiphase synchronization for
co-rotating orbits \cite{LeoniLiverpool}.   The path to solving this mystery will likely involve the continued use of mutants, genetics, and
micromanipulation \cite{eLife} to isolate the
crucial features of each mode of synchronization. 

The above discussion of synchronization has neglected any detailed description of the flagellar waveform itself, 
but in many ways this is one of the most important outstanding problems in the field.  Although there is consensus that the
undulations are ultimately a consequence of microtubule doublets being slid past one another by the stepping of
dynein motor proteins, the manner in which this translates into the observed waveform is still under debate 
\cite{Mitchison_Nature}.  One promising approach \cite{RiedelKruse} builds on the idea of spontaneous
oscillations of active filaments \cite{Brokaw,Camalet}.  Here, the coupling among motor stepping, probabilistic
detachment, filament bending and viscous resistance leads to an eigenvalue problem for a Hopf bifurcation.  This
immediately shows that there is a discrete set of modes on the eukaryotic flagellum, each with a distinct waveform
and frequency.  Perhaps the wild-type breaststroke mode and the {\it ptx1} antiphase mode are the first two rungs on
this ladder.  Intriguingly, {\it Chlamydomonas} is known to exhibit a photoshock response in which
the two flagella undulate in the sinusoidal shape mode in front of the cell body, propelling it backwards away from
intense light \cite{photoshock}.  This beating occurs at yet higher frequencies.  Determining whether this picture of discrete modes 
holds is an important goal
for the next generation of experiments.

\subsection{Multitudes}

Some of the most important situations in which cilia and flagella drive fluid motion involve great numbers of them
emanating from tissue.  This is the case in the left-right organizer (LRO) that is involved in embryonic symmetry breaking
\cite{LRO}, in the respiratory tract and reproductive system of vertebrates, and in ciliates such as
{\it Paramecium} and {\it Opalina}.   The nodal cilia in the left-right organizer are short structures that whirl around, but
in all the other examples we find longer cilia with well-defined power and recovery strokes, exhibiting metachronal waves.  Despite 
decades of observation, there
has been remarkably little quantitative study of the dynamics and fluid dynamics of metachronal waves \cite{KnightJones}.  
This is perhaps not surprising for the
animal contexts, given the difficulty of visualization {\it in vivo} or even {\it ex vivo}.   {\it Volvox} has again provided a way forward,
for close observation has demonstrated that it displays robust metachronal waves \cite{Metachronal}.  Figure \ref{fig7}c shows
a space-time illustration of one component of the fluid velocity of the metachronal waves of {\it Volvox carteri} just above the flagella tips,
along with various statistical measures of the metachronal wave properties.    
A typical wavelength is on the order of $10$ somatic cells,
so the relative phase shift between neighbors is small.  

One way to think about metachronal waves is to compare the behavior of pairs of flagella.
If the basic coupling between {\it Chlamydomonas} flagella leads to precise phase locking, why does a carpet of
flagella display long-wavelength modulations in phase?  Is this a cumulative effect of interactions with multiple
flagella?  Does it arise from the presence of an underlying no-slip surface such as the extracellular matrix of
{\it Volvox} or a tissue layer in the respiratory system?  Although there are theoretical studies that take as a starting point the
existence of metachronal waves and examine their consequences for swimming \cite{Blake_metachronal,Brennen_metachronal,Lauga_metachronal},
only more recently have detailed proposals for the underlying mechanism been advanced.  These include analyses of undulating filaments near
no-slip surfaces \cite{Guirao_metachronal,GuiraoPCP} and coupled orbiting spheres near surfaces \cite{Julicher_metachronal,Metachronal}.
In each case the common feature is the presence of the no-slip boundary, suggesting that the deflection of the flow by it produces the observed phase shift.

\section{PHOTOTAXIS}
\label{phototaxis}

As photosynthetic organisms, the volvocine algae need light for survival, and evolution has endowed them with
photoreceptors termed eyespots to enable steering toward the light \cite{Jekelyreview}. {\it Chlamydomonas} has a single eyespot, as does
each somatic cell of multicellular species such as {\it Volvox}.  In the latter case, the eyespots tend to have a gradation
in size, from large ($\sim 2.5$ $\mu$m in diameter) at the anterior to small ($\sim 0.8$ $\mu$m) near the posterior \cite{Fidelity}.  
In the volvocine algae these eyespots have a distinctive reddish color.
Each eyespot lies inside the chloroplast of the cell and has a remarkable internal structure comprising multiple layers of
proteins and membranes that function as a quarter-wave plate \cite{Foster}.  The result is a structure which is sensitive to light
impinging on it in the hemisphere around the outward surface normal but insensitive to light from behind.  This
is a crucial property for the mechanism of phototactic steering. 

\begin{figure*}[t]
    \begin{center}
        \includegraphics*[clip=true, width = 1.9\columnwidth]{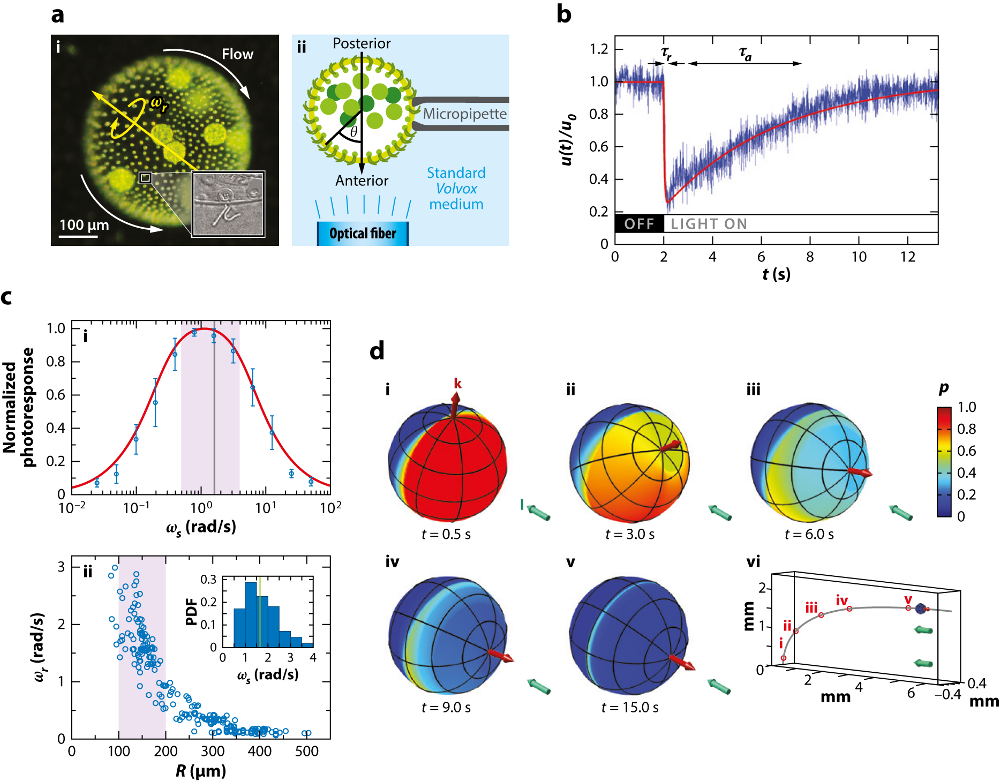}
    \end{center}
    \caption{Phototaxis in {\it Volvox carteri}. ({\it a}) Experimental setup. ({\it i}) Flows around the colony. ({\it ii}) 
Optical fiber illumination setup. 
({\it b}) Local flagella-generated fluid speed $u(t)$ ({\it blue}), measured with particle image velocimetry just above the flagella 
during a step-up 
in light intensity, normalized by the flow speed in the dark. Two timescales are evident: a short response time, $\tau_r$, and a 
longer adaptation time, 
$\tau_a$. The fitted theoretical curve ({\it red}) is from the solution of Equations 10 and 11. ({\it c}) Photoresponse frequency 
dependence and colony rotation. 
({\it i}) Normalized flagellar photoresponse versus frequencies of sinusoidal stimulation ({\it blue circles}). The theoretical 
response function 
({\it red line}; Equation 12) shows quantitative agreement. ({\it ii}) Colony rotation frequency $\omega_r$ of {\it V. carteri} as a 
function of radius $R$. 
Highly phototactic organisms within the range indicated by the purple shaded region, corresponding to the purple shaded region 
in part {\it i}, 
demonstrate that response timescales and the colony rotation frequency are mutually optimized to maximize the photoresponse. 
({\it d}) Colony 
behavior during a phototurn. ({\it i--v}) The colony axis ({\it red arrow}) tipping toward the light direction I ({\it aqua arrow}). 
Colors represent 
the amplitude $p(t)$ of the downregulation of flagellar beating in a model of phototaxis. ({\it vi}) The swimming trajectory. 
Abbreviation: 
PDF, probability distribution function. Figure adapted with permission from Drescher et al. (2010b). Copyright 2010 by the 
National Academy of Sciences.}
    \label{fig8}
\end{figure*}

As noted above, {\it Volvox} was given its name for its spinning motion, the frequency of which ranges from $\sim 0.3$ Hz for
small, young colonies of {\it V. carteri} to $0.02$ Hz for larger, older colonies.  Those that are very phototactically active
tend to have frequencies toward the upper range.  {\it Chlamydomonas} also spins about its major axis as it swims, typically
at frequencies of $\sim 2$ Hz.  When coupled with the directionality of the eyespot, these motions give rise to light 
intensities at the eyespots that are periodic functions of time when the organism (single cell or colony) is not aligned with the
incident light and that are nonmodulated signals when it is aligned.  It appears therefore that nature has repeatedly found a
biochemical mechanism to use periodic signals as a means to navigate in the sense that the existence of periodicity implies
a lack of proper alignment.  There are two clear questions that arise from these basic observations.  First, what is the
relationship between the light signals and flagellar beating?  Second, how does this result in navigation 
toward light?  It should be noted that the interplay of helical swimming and periodic signals appears to be more
general than phototaxis, as there is evidence that sperm {\it chemotaxis} may operate similarly \cite{FriedrichJulicher_spermchemo}.  
The ubiquity of spiral swimming has been remarked upon for over a century
\cite{Jennings}.

It is intuitive that, in the absence of light stimulus, an organism swimming along a straight trajectory through
the symmetric action of its flagella can only 
deviate from that path by an asymmetry in its flagellar beating.  In the case of {\it Chlamydomonas}, the synchronization
described in earlier sections occurs despite the intrinsic differences between the two flagella.  Not only are the intrinsic
beat frequencies distinct, but so too is their response to intracellular calcium levels.
This has been investigated in great detail using so-called cell models, which are cells that have been demembranated yet 
still function.  Without the membrane, it is possible for the experimentalist to alter the ambient concentration of calcium ions 
experienced by the flagella.  The two flagella respond differently to calcium, so that the {\it cis} flagellum is the dominant
one at low calcium, and {\it trans} dominates at high levels. This property is known as
flagellar dominance \cite{flagellar_dominance}.  {\it In vivo}, flagellar dominance is crucial, for when 
light falls on the eyespot, there is a transient change in the
calcium levels within the cell (owing to the opening of membrane-bound calcium channels), leading to transient changes in the
frequencies or amplitudes of the beating of the two flagella.  The flagella transiently fall out of precise synchrony and produce a turn. 
{\it Chlamydomonas} swims in helical trajectories with its eyespot facing outwards, and there must be
very precise tuning of the transient asymmetries to enable phototaxis \cite{Schaller}.  Indeed, \textcite{Yoshimura} found
a tuning of the flagellar response dynamics to the spinning frequency of the colony, a feature detailed below for {\it Volvox}.

In the case of a multicellular organism such {\it Volvox}, the problem is even more interesting.  First, as the two
flagella on the somatic cells beat in the same direction, as opposed to the breaststroke geometry of {\it Chlamydomonas},
steering of the colony as a whole cannot in any obvious way arise from differential flagella beating \cite{Hoops}. Second,
as demonstrated long ago \cite{HiattHand}, there is no difference in phototactic ability between species of
{\it Volvox} that have cytoplasmic connections between somatic cells in the adult and those that do not.  Hence, there
is no need for cytoplasmic cell-cell communication for phototaxis to operate.
As there is no central nervous system in {\it Volvox} it is apparent that if the organism can
steer to light, then it must be because each somatic cell responds appropriately by itself.  The appearance of coordination
in the response is illusory!  The question is thus: What is the right program of response to a periodic light signal?

Ever since the work of \textcite{Holmes}, it has been clear that some sort of asymmetry in the flagellar beating 
between the light and dark sides of a colony is necessary for phototactic turning, but only recently has a quantitative
theory been proposed and validated by two groups, independently and nearly simultaneously \cite{Fidelity,Ueki}.
Many issues arising in these studies also appear in an important earlier work on the mechanism of phototaxis of
a marine zooplankton \cite{JekelyNature}.
Although ultimately it is desirable to understand the phototactic response at the level of the individual somatic cells and their
flagella, a useful coarse-grained approach builds on the important work of \textcite{StoneSamuel}, in which the
angular velocity ${\mathbf{\Omega}}$ of a spherical swimmer at low Reynolds numbers is expressed in 
terms of the fluid velocity field ${\bf u}$ on its
surface:
\begin{equation}
{\mathbf{\Omega}} (t) = -\frac{3}{8 \pi R^{3}}\int \hat{\mathbf{n}}\times\mathbf{u}(\theta,\phi,t) \:
\mbox{d}S,
\label{eq:phototurn}
\end{equation}
where $\hat{\mathbf{n}}$ is the local normal to the sphere.  The need for non-uniformity in ${\bf u}$ for turning
of the colony axis is readily apparent from this relationship.  

Adopting this coarse-grained approach, one can utilize
time-resolved PIV to measure the response of a carpet of flagella on the {\it Volvox} surface to changing light
conditions.  Rather than dealing with freely swimming colonies, it is far simpler to hold them on micropipettes and
mimic the effects of colony rotation by turning on and off illumination light at particular frequencies.  For confirmation
of the directionality, such light is sent through a fiber-optic light guide held within a micropipette near the colony
surface (Figure \ref{fig8}a).  The essential result of this investigation is shown in Figure \ref{fig8}b, which displays the transient response of the
local fluid flow to a sudden illumination of a patch on the surface, mimicking the effects of colony rotation in bringing that
patch into the light.  There is a rapid ($\sim 50$ ms) decrease in the fluid flow, followed by a much slower
recovery back to the original level.  This is an adaptive response in the same sense that our eyes or olfactory system
adjusts to the sudden onset of a strong signal.
If $u_0$ is the local fluid speed before such perturbations, then we find $u(t)/u_0= 1-\beta p(t)$, where $p(t)$ is a dimensionless 
photoresponse variable.  A dynamical model for $p(t)$ that captures the experimental behavior invokes the 
existence of a second, hidden variable $h(t)$, which represents the internal biochemical mechanism of adaptation.
These are coupled together in a manner already known from work on bacterial chemotaxis, in which adaptation is
well understood:
\begin{eqnarray} 
\tau_r\dot{p}&=&(s-h) \: \: H (s-h) - p ~, \label{eq:adapt1} \\
\tau_a\dot{h}&=&s - h   ~, \label{eq:adapt2}
\end{eqnarray}
where $s(t)$ measures the light stimulus; $\tau_r$ and $\tau_a$ are the response and adaptation timescales, respectively;
and the  Heaviside function ensures that a step down in light stimulus cannot increase
$u$ above $u_0$.  This pair of ODEs has a stable fixed point at a constant stimulus, $h$, and $p=0$, and will display
precisely the biphasic response seen in Figure \ref{fig8}b under step changes in $s$.  

The key implication of this model is that the frequency response of the coupled system exhibits a resonance, a feature also known
from studies of {\it Chlamydomonas} \cite{Yoshimura}.  This can be
seen by Fourier transforming Equation \ref{eq:adapt2} (neglecting the Heaviside function for simplicity), yielding
\begin{equation}
{\cal R}(\omega_s)=\frac{\omega_{s} \tau_a}{\sqrt{(1+\omega^{2}_s
\tau^{2}_r) (1+\omega^{2}_s \tau^{2}_a)}}~. 
\label{eq:responsivity}
\end{equation}
This demonstrates immediately that if the frequency of the stimulus is either too large or too small the response is small.
Most interestingly, the peak of the response (Figure \ref{fig8}c) occurs for $0.4 < \omega < 4$ rad/s, which match
the orbital frequencies of those colonies that have the strongest phototactic response.   The effect of this tuning is
clear: If a patch on the surface rotates into the light, its flagellar beating is downregulated, and if the recovery from that
state takes a time on the order of a rotation period, the net effect is that the illuminated side of the colony has weaker
beating and the dark side has stronger beating.  This is the asymmetry needed.  It follows that if one can artificially slow
down colony rotation while keeping the adaptive response unchanged, then the phototactic ability should decrease.  Experiments
verify this directly \cite{Fidelity}.  The analysis is completed by solving for the trajectories of swimming colonies using the adaptive model
over the entire surface.  The result clearly illustrates how the initial asymmetries from misalignment with the light are
diminished as the colony turns toward the light, with a stable fixed point at perfect alignment  (Figure \ref{fig8}d).

\section{TRACER STATISTICS IN SUSPENSIONS}
\label{tracers}

Interest in the properties of concentrated suspensions of microorganisms has expanded greatly since early suggestions
\cite{Vicsek,TonerTu} of the possibility that self-propelled organisms could display a transition to long-range order.  
Although subsequent experimental work  showed that such hypothesized order does not occur \cite{WuLibchaber,Dombrowski},
the observed dynamics is in many ways more interesting.  It takes the form of transient, recurring vortices and jets of
coherent swimming with characteristic length scales reaching $100$ $\mu$m and beyond, far larger than the individual swimmers.  
There is a continuing debate in the literature regarding the mechanism underlying this
behavior - whether it arises purely from steric effects or requires long-range hydrodynamic interactions.  This is reviewed
in detail elsewhere.  Here, the focus is on the issue of how these turbulent flows \cite{Bacterial_turbulence} 
impact the transport of 
suspended tracer particles.  From the
pioneering work of \textcite{WuLibchaber} and later work by others \cite{Angelani} 
swimming bacteria can be viewed as analogous
to molecules in a conventional fluid whose collisions with tracers, such as the pollen grains of \textcite{Brown}, 
produce Brownian motion.  Whereas in conventional Brownian motion with a molecular bath and micrometer-sized particles there is
an enormous separation of timescales between the picoseconds of molecular collisions and the perhaps milliseconds on which we
resolve the tracer displacements, no such separation exists in microorganism suspensions.  Instead, we are able to resolve the 
collisions of the swimmers with the tracers.  

\begin{figure*}[t]
    \begin{center}
        \includegraphics*[clip=true, width = 1.8\columnwidth]{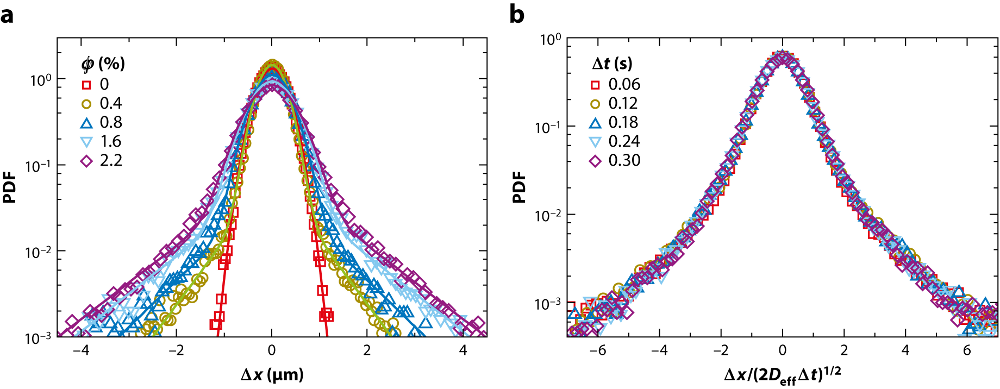}
    \end{center}
    \caption{Probability distribution functions (PDFs) for tracer displacements in suspensions of {\it Chlamydomonas}. 
({\it a}) PDFs at a fixed time interval $\Delta t=0.12$ s, at various volume fractions $\phi$. The Gaussian develops apparently 
exponential tails as $\phi$ increases. (b) Diffusive rescaling of PDFs at $\phi=2.2$\% for several time intervals ($\Delta t$), 
leading to data collapse to a non-Gaussian function. Figure adapted with permission from Leptos et al. (2009). 
Copyright 2009 by the American Physical Society.}
    \label{fig9}
\end{figure*}

The first and most basic observation from studies of bacterial suspensions is that the tracer particles exhibit enhanced
diffusion with increasing swimmer concentration, with diffusivities reaching $\sim 10^{-6}$ cm$^2$/s at cell
concentrations of $n\sim 5\times 10^{10}$/cm$^3$, orders of magnitude larger than the purely thermal values of $D$.
These measurements were obtained by tracking of the trajectories of individual tracers and calculating the 
mean-square displacement $\langle \Delta r^2\rangle$ versus time (t).  Such analyses show at short times a scaling
$\langle \Delta r^2\rangle\simeq t^{\alpha}$ with $1.5 < \alpha <2$, not quite the ballistic scaling $\alpha=2$ one might expect,
but definitely superdiffusive.  At larger times ($t\gtrsim 3$ s) there is a clear crossover to diffusive behavior, with $\alpha=1$
holding beyond.  Of course, the mean-square displacement is only one moment of a distribution of particle displacements observable
over some time interval $t$, and more details are available in the full PDF.
In this case, {\it Chlamydomonas} has a number of advantages over the use of bacteria.  First, its mean swimming speed is considerably
larger than that of typical bacteria, so one would anticipate stronger effects on the tracer particles.  Second, the organism is
larger than easily observed tracer particles, and is therefore likely to be unaffected by their presence.  Moreover, 
it is possible to visualize the tracers' encounters with the swimmers in detail.  This is then not unlike the situation
considered  by \textcite{Darwin} in his analysis of the trajectories of particles in an {\it inviscid} fluid as a body
is moved through the fluid, giving rise to the so-called Darwin drift.

Figure \ref{fig9}a shows the results from the study of tracer particles in {\it Chlamydomonas} suspensions \cite{TracerDiffusionPRL} 
as a semilogarithmic 
graph of the PDF of particle displacements $\Delta x$ along one axis of observation, at a
fixed time interval $\Delta t$ for various volume fractions $\phi$ of swimmers.  For $\phi=0$, we see an inverted parabola, 
signifying a familiar Gaussian PDF, $P(\Delta x)\sim \exp(-\Delta x^2/2D\Delta t)$, where $D$ is the Stokes-Einstein
diffusivity of the tracers.  At higher volume fractions, the distribution develops a heavy tail, consistent with an exponential
form.  One infers that the core of each of these distributions continues to arise from the thermal Brownian motion of the tracers, 
whereas the tails result from the fluid flows created by the swimmers. 
Strikingly, as shown in Figure \ref{fig9}b, a diffusive rescaling collapses all the data at various $\Delta t$ at a given $\phi$.
At almost exactly the same time as this discovery, Granik and collaborators found analogous behavior in two
completely different systems, and dubbed it `anomalous but Brownian'\cite{GranikPNAS}.  
Their systems were (a) colloidal microspheres that diffuse 
along phospholipid bilayer tubes of the same radius and (b) microspheres diffusing through networks of filamentous actin,
in which the mesh size is considerably larger than the spheres.  Although the mean-square displacement of the spheres in both cases
grew linearly with time, the PDFs were decidedly non-Gaussian.  Later work \cite{Gollub2D} on tracer dynamics in {\it Chlamydomonas} 
suspensions confined to thin 
films revealed even stronger deviations from Gaussianity than in three dimensions \cite{TracerDiffusionPRL}.  
In the three-dimensional case, an examination of the trajectories of particles near 
the alga reveals complex loops qualitatively similar to those of the Darwin problem,  
consistent with recent calculations of tracer paths at zero Reynolds number 
\cite{DunkelPutz,JornYeomans,PushkinJFM}.

The notion that these non-Gaussian distributions are anomalous arises from the expectation that the random encounters between
swimmers and tracers would satisfy the conditions of the central limit theorem (bounded second moment of the distribution).
But things are not so simple because of the long-range flow fields around the swimmers \cite{ProtistFluctuations}.
If we have a suspension of $N$ swimmers, each of radius $R$, in a box of linear dimension $L$, and consider a volume fraction
$\phi=4\pi R^3N/3L^3$ that is sufficiently small, the PDF of velocities 
arises from a random superposition of the flow fields 
around each swimmer.  Averaging over their positions is equivalent to integrating over space when the distribution is uniform. 
Let us suppose  that the velocity around a swimmer
decays as $\vert{\bf v}(r)\vert\sim A(\Omega)/r^n$, with $\Omega$ an angular factor.  Then the probability
distribution $P(v)$ of velocities is
\begin{equation}
\label{eq1}
P(v)\propto L^{-3}\int_0^L\int_{D_{\Omega}}\delta\Bigl(v-\frac{A(\Omega)}{r^{n}}\Bigr)r^2 dr d\Omega ~.
\end{equation}
A simple scaling argument shows that the tail of $P$ has the form
\begin{equation}
P(v)\propto\frac{1}{v^{1+3/n}}.
\label{scaling}
\end{equation}
The second moment of $P(v)$ is finite only if $n<3/2$, the case of a 
Stokeslet ($n=1$). This is the condition for the validity of the central limit theorem; the
velocity field from a large number of independently placed Stokeslets 
is Gaussian. It will not be so for any higher integer singularity, such as stresslets ($n=2$) 
or source doublets  ($n=3$) \cite{BlakeChwang74}.  
If the decay law deviates from $v\propto r^{-n}$ below a certain radius, 
the PDF shape (Equation \ref{scaling}) will break down above the corresponding value of $v$.  Hence, 
deviations from Gaussianity provide a direct probe of the near-field velocity around the swimmers.  Experiments on velocity fluctuations in
suspensions of {\it Volvox} explicitly demonstrated that the presence or absence of the Stokeslet contribution has a strong effect on
the fluctuation spectrum \cite{ProtistFluctuations}.

Several explicit models have been proposed to connect the near-field behavior to the non-Gaussian statistics. 
\textcite{Thiffeault} used a squirmer calculation to examine the displacements of tracers in the path of a swimmer and were able to find non-Gaussian 
statistics arising from the proximity to a stagnation point on the squirmer surface, although the non-Gaussianity had different scaling properties from 
those seen
in experiment.  In related work, \textcite{Eckhardt_stirring} examined the implications of power-law distributions of trapping times and step lengths in a 
continuous-time random 
walk \cite{JornYeomans} and found conditions that gave good agreement with experiment.  It was suggested that these 
distributions could arise from the existence of 
stagnation points in the flow.   These are promising developments in understanding the non-Gaussianity seen in experiments.  Future 
experiments that can visualize which tracer displacements give rise to the heavy tails can provide tests of these theories.

\section{SUMMARY POINTS}

1.   The average and time-resolved flow fields around individual freely swimming microorganisms have been measured and interpreted in terms of
elementary force singularities.  Attention should now focus on connecting those velocity fields to the detailed action of the underlying flagella.

2.   Interactions of microswimmers with boundaries can induce striking dynamics.  Some can be explained by purely hydrodynamics mechanisms,
but others require explicit consideration of the contact interactions between the flagella and the surface.

3.  There is strong evidence that the synchronization of eukaryotic flagella can occur through 
hydrodynamic interactions, but the particular
mechanism is still under investigation.   The synchronization dynamics is intrinsically stochastic and subject to intracellular biochemical changes on
intermediate timescales.  Mutants have revealed unusual modes of synchronization that provide a challenge to current theories.

4.  The mechanism of phototaxis in multicellular green algae has been shown to involve a tuning between the adaptive response time of the
flagellar apparatus to changing light levels and the orbital period of the spinning organisms.  This mechanism does not require any 
explicit communication between cells in the colony.

5.   Suspensions of swimming algae can exhibit non-Gaussian yet Brownian statistics of tracer particle displacements.  These are thought to 
arise from the detailed form of the flow field around the swimmers.

\section{FUTURE ISSUES}

1.  The demonstration of the {\it in vitro} evolution of multicellularity, first with yeast \cite{TravisanoPNAS} and then with
{\it Chlamydomonas} \cite{TravisanoNature}, has shown that selection based on simple hydrodynamic properties (e.g., 
gravitational settling speed) can induce
multicellularity from organisms previously thought of as strictly unicellular.   
What would happen with other selective pressures?  What can theory say about these results?  

2.  The recent discovery \cite{SmithB12} of a symbiotic relationship between bacteria and algae, in which bacteria provide needed 
vitamin B$_{12}$ to the algae, raises a host of fascinating questions in physical ecology.  How do these organisms find each 
other in the vastness of the oceans?  How does the symbiosis persist in the presence of turbulence?  More generally,
what are the spatiotemporal dynamics of such symbioses?

3.  A quantitative theory for phototaxis in {\it Chlamydomonas} is 
still lacking, as are tests of the tuning mechanism for species intermediate between {\it Chlamydomonas} and {\it Volvox}.
The origin of the slow timescale in the adaptive phototactic response of {\it Volvox} is unclear from a biochemical
perspective, yet it is crucial for the tuning mechanism. 

4.  The existence of both in-phase and antiphase synchronization in the {\it ptx1} mutant of {\it Chlamydomonas} 
remains unexplained at a mechanistic level.  Similarly, a test of the hypothesis that these are two among many possible
discrete undulation modes of the eukaryotic flagellum is needed.

5.  A detailed understanding of metachronal waves is still lacking, as are quantifications of the stochastic dynamics
of ciliary carpets on large length scales and timescales.  {\it Volvox} may be the ideal organism for such studies.

\newpage

\noindent{\bf Disclosure statement}

\smallskip

The author is not aware of any biases that might be perceived as affecting the objectivity of this review.

\medskip

\noindent{\bf Acknowledgments}

\smallskip

I am deeply indebted to Cristian Solari, John Kessler, and Rick Michod for first interesting me in the evolution of 
multicellularity and the Volvocales, and I thank them and D. Brumley, K. Drescher, J. Dunkel, S. Ganguly, J.S. Guasto, J.P. Gollub, 
M. Herron, V. Kantsler, K.C. Leptos, T.J. Pedley, A.I. Pesci, M. Polin, M.B. Short, I. Tuval, and K. Wan for collaborations.  
This work was supported by NSF grant PHY-0551742, BBSRC grant BB/F021844/1, ERC Advanced Investigator Grant 247333, 
the Leverhulme Trust, the Schlumberger Chair Fund, and a Wellcome Trust Senior Investigator Award.

\bibliography{Goldstein_bib}

\end{document}